\documentclass[preprints,review,accept,moreauthors,pdftex]{mdpi}
\firstpage{1} 
\pubvolume{1}
\issuenum{1}
\articlenumber{0}
\pubyear{2022}
\copyrightyear{2022}
\datereceived{2022-01-07} 
\dateaccepted{} 
\datepublished{} 
\hreflink{https://doi.org/} 

\usepackage[export]{adjustbox}
\usepackage{supertabular}

\Title{Gamma-ray Cosmology and Tests of Fundamental Physics}

\TitleCitation{Gamma-ray Cosmology and Tests of Fundamental Physics}

\Author{Jonathan Biteau $^{1}$\orcidA{} and Manuel Meyer $^{2}$\orcidB{}}

\AuthorNames{Jonathan Biteau and Manuel Meyer}

\AuthorCitation{Biteau, J.; Meyer, M.}

\address{%
$^{1}$ \quad Universit\'e Paris-Saclay, CNRS/IN2P3, IJCLab, 91405 Orsay, France;\\
$^{2}$ \quad Institute for Experimental Physics, University of Hamburg, Luruper Chaussee 149, 22761 Hamburg, Germany;}

\corres{Correspondence: biteau@in2p3.fr, manuel.meyer@uni-hamburg.de.}

\abstract{The propagation of gamma-rays over cosmological distances is the subject of extensive theoretical and observational research at GeV and TeV energies. The mean free path of gamma-rays in the cosmic web is limited above 100 GeV due to the production of electrons and positrons on the cosmic optical and infrared backgrounds. Electrons and positrons cool in the intergalactic medium while gyrating in its magnetic fields, which could cause either its global heating or the production of lower-energy secondary gamma-rays. The energy distribution of gamma-rays surviving the cosmological journey carries observed absorption features that gauge the emissivity of baryonic matter over cosmic time, constrain the distance scale of $\Lambda$CDM cosmology, and limit the alterations of the interaction cross section. Competitive constraints are in particular placed on the cosmic star-formation history as well as on phenomena expected from quantum gravity and string theory, such as the coupling to hypothetical axion-like particles or the violation of Lorentz invariance. Recent theoretical and observational advances offer a glimpse of the multi-wavelength and multi-messenger path that the new generation of gamma-ray observatories is about to open.}


\begin{document}

\section{The extragalactic gamma-ray sky}

The total electromagnetic emission from astrophysical objects in the Universe is dominated by thermal emission from stars, gas and dust. Their optical and infrared emission, originating from nuclear processes and thermal radiation, dominates the cosmic background radiation at wavelengths ranging from 0.1\:$\mu$m to 100\:$\mu$m \cite{2018ApSpe..72..663H}. These optical and infrared backgrounds are collectively referred to as the extragalactic background light (EBL). While millions of galaxies have been observed over the full-sky in the optical band, only a fraction of them also emits non-thermal radiation at radio and X-ray wavelengths \cite{2017A&ARv..25....2P}. Such non-thermal emission originates from charged particles, electrons and positrons or protons and nuclei, which are accelerated by violent processes in astrophysical environments. These charged particles radiate away their energy by gyrating in magnetic fields or by interacting with particles and lower-energy photons populating the surrounding medium \cite{2013LNP...873.....G}. The frontier of the electromagnetic spectrum up to which such astrophysical accelerators are observed is the gamma-ray domain, which is defined as the electromagnetic band covering energies larger than a few MeV. 

Gamma-rays from astrophysical sources are observed with dedicated particle-physics instruments either in space or on the ground. The most successful technique at energies from 100\:MeV to 1\:TeV is based on detectors orbiting Earth, such as \textit{Fermi}-LAT \cite{2009ApJ...697.1071A}, that combine a tracker and a calorimeter in which gamma-rays generate showers of electrons and positrons. The hits and charge deposits from the secondary leptons are extracted to reconstruct the arrival direction and energy of the primary gamma-ray. The effective area of such space detectors, on the order of a square meter, is a limiting factor beyond few hundreds of GeV. The non-thermal emission from astrophysical sources follows to first order a power-law spectrum of energy, with a differential number of photons per infinitesimal energy band that strongly decreases with energy, $E$, typically as $E^{-2}$. In the very-high-energy range, from 100\:GeV to 100\:TeV, the atmosphere of Earth itself is used as a calorimeter, so that ground-based telescopes separated by one hundred meters, such as H.E.S.S., MAGIC and VERITAS, build up an effective collection area ten thousand times larger than that of gamma-ray satellites \cite{2009ARA&A..47..523H}. 

The field of high-energy astrophysics, dedicated to the study of non-thermal emission, is relatively young compared to thermal astrophysics (see, e.g., Refs.~\cite{2015CRPhy..16..587D, 2016CRPhy..17..594D} and references therein for a historical perspective). The development of the radar technique during World War II enabled in the 1950s the first detections of radio galaxies such as Centaurus\:A (at a luminosity distance $d \sim 4$\:Mpc), M\:87 ($d \sim 16$\:Mpc) and Cygnus\:A ($d \sim 230$\:Mpc). Such radio observations, combined with optical spectroscopy, triggered the development of multi-wavelength extragalactic astronomy in the early 1960s, with the first measurements of redshifts of quasars at cosmological distances, namely 3C\:273 and 3C\:48 at redshifts $z=0.16$ and $z=0.37$, i.e.\ $d \sim 770$\:Mpc and $d \sim 2000$\:Mpc, respectively \cite{1990mcr..book..347S}. The late 1960s and early 1970s revealed that some of these radio galaxies and quasars, such as M\:87, Centaurus\:A and 3C\:273, also emit X rays. The quasar 3C\:273 was first detected in the gamma-ray band in 1978 by the CoS-B satellite \cite{1978Natur.275..298S} and remained the only known extragalactic gamma-ray source for over a decade. The non-thermal luminosity of such quasars, nowadays encompassed together with radio galaxies in the more general class of active galactic nuclei (AGNs \cite{2017A&ARv..25....2P}), goes up to $\sim\: 10^{41}\:$W which is equivalent to hundreds of trillions of solar luminosities, that is thousands of times larger than the stellar emission from the host galaxy. The tremendous luminosities of AGNs like 3C\:373 are enabled by the relativistic beaming and energy shift of photon fields from jetted outflows. If the jet is closely aligned with the line of sight, the AGN is called a blazar. The observed photon energy is then enhanced with respect to the emitted photon energy in the jet frame, by a Doppler factor $\delta \sim 10$. The observed bolometric luminosity is further enhanced by a factor $\delta^4$ \cite{2013LNP...873.....G}. The gamma-ray band is of particular interest in the study of jetted AGN for two reasons: not only does it probe the highest energies per photon but it also often encompasses more than half of the electromagnetic luminosity of AGNs, thus providing an intense and energetic beam to study cosmological processes and fundamental physics. 

\begin{figure}[t!]
\includegraphics[width=1\linewidth]{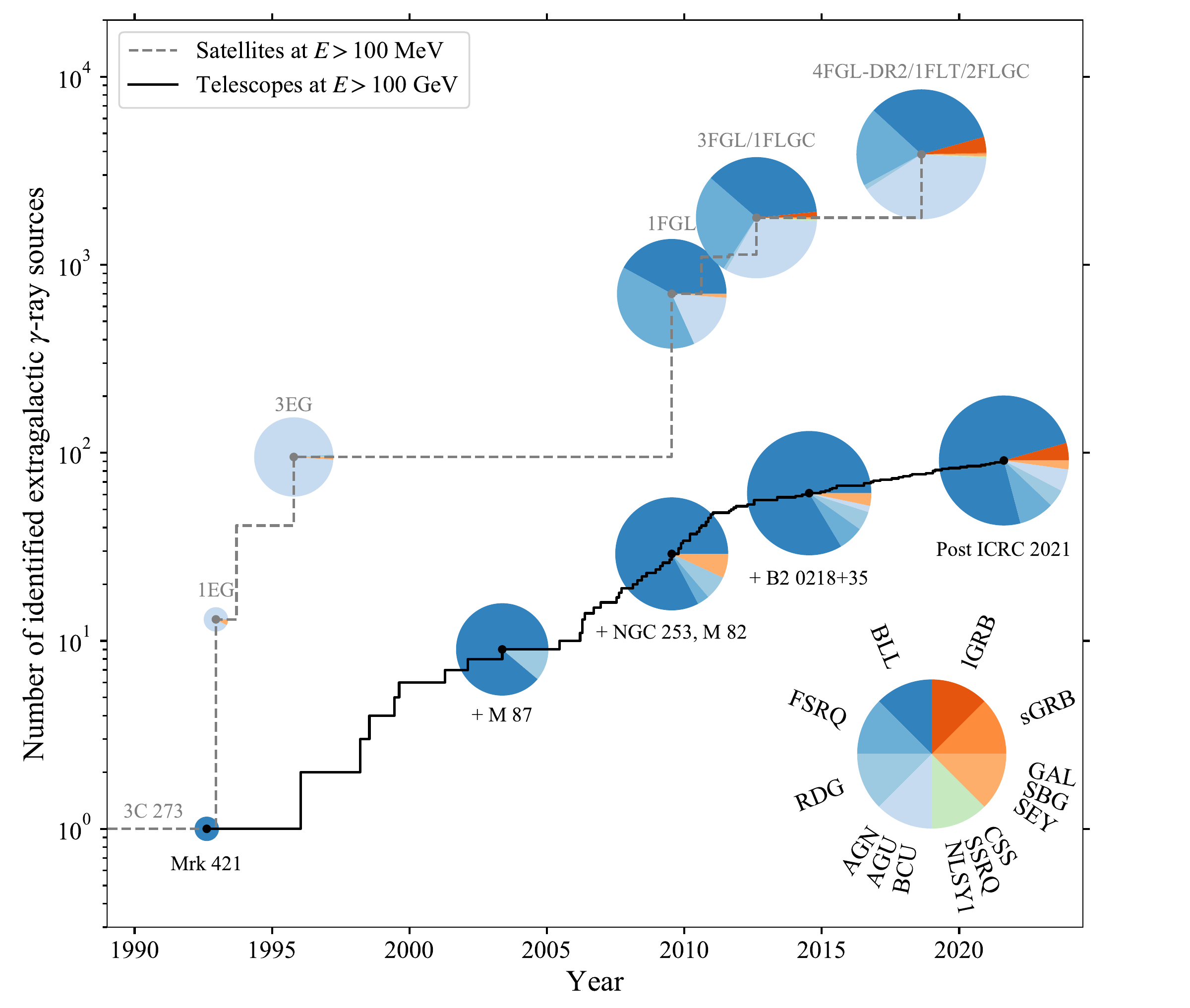}
\caption{Evolution of the number of extragalactic gamma-ray sources at high energies as a function of time. The extragalactic sources are usually identified as such by matching them with multiwavelength counterparts. The dashed gray line shows significant detections by the EGRET and \textit{Fermi}-LAT gamma-ray satellites at energies larger than 100\:MeV. The pie charts display the distribution of sub-classes of extragalactic sources, as extracted from the 1EG \cite{1994ApJS...94..551F}, 3EG \cite{1999ApJS..123...79H}, 1FGL \cite{2010ApJS..188..405A}, 3FGL \cite{2015ApJS..218...23A}, 1FLGC \cite{2013ApJS..209...11A}, 4FGL-DR2 \cite{2020ApJS..247...33A}, 1FLT \cite{2021ApJS..256...13B} and 2FLGC \cite{2019ApJ...878...52A} catalogs. The solid black line shows significant detections by gamma-ray telescopes above 100\:GeV as extracted from TeVCat \citep{2008ICRC....3.1341W} after the 2021 international cosmic-ray conference (ICRC 2021). Most of the pre-2005 discoveries were performed at the Whipple observatory while later discoveries essentially came from H.E.S.S., MAGIC and VERITAS. Some of the emblematic ground-based discoveries are indicated under landmark pie charts, featuring the detections of the first blazar Mrk\:421 (Whipple), radio galaxy M\:87 (HEGRA), starburst galaxies NGC\:253 (H.E.S.S.) and M\:82 (VERITAS), and gravitationally-lensed blazar B2\:0218+35 (MAGIC). The bottom-right pie chart illustrates the categories in which sources are grouped. In blue: BL\:Lac (BLL: 1307/3866 and 68/91 at $E>100\:$MeV and $E>100\:$GeV, respectively, at the time of writing) and flat-spectrum radio quasar (FSRQ: 762/3866 and 8/91) for blazars, radio galaxy (RDG: 46/3866 and 4/91) as their non-aligned counterpart, active galactic nuclei of either radio galaxy of blazar-type with uncertain classification (AGN, AGU, BCU: 1545/3866 and 5/91); in green: the emerging gamma-ray class of narrow-line Seyfert 1, steep-spectrum radio quasars and compact steep spectrum sources (NLSY1, SSRQ, CSS: 17/3866 and 0/91); in orange: regular star-forming galaxies together with starburst galaxies and non-jetted Seyfert-type AGN (GAL, SBG, SEY: 20/3866 and 2/91), gamma-ray bursts of short (sGRB: 14/3866 and 0/91) and long (lGRB: 155/3866 and 4/91) duration.\label{fig:kifune_xgal}}
\end{figure}

Extragalactic gamma-ray astronomy was established as a disciplinary field in its own right during the 1990s, with the successful operation of the Whipple telescope on Mount Hopkins (Arizona, USA) and the launch of EGRET on board the CGRO satellite. Figure~\ref{fig:kifune_xgal} illustrates the growth of the number of gamma-ray sources identified outside of the Milky Way since that epoch.  At the turn of the millennium, space-borne EGRET observations had enabled the firm identification over the entire sky of nearly a hundred extragalactic sources above 100\:MeV (see 3EG Catalog \cite{1999ApJS..123...79H}). Cross-matches with lower-energy multiwavelength observations revealed that most of these accelerators are AGN, although without specific sub-classification (see AGN, AGU, BCU in 3EG pie chart of Figure~\ref{fig:kifune_xgal}). The only non-AGN extragalactic source was the Large Magellanic Cloud, which is the Milky Way's largest satellite galaxy at $d \sim 0.05\:$Mpc. Meanwhile, the first ground-based gamma-ray telescopes, with their limited field of view of a few degrees, were pointed to the most promising targets selected among radio, optical and X-ray extragalactic sources. The first extragalactic gamma-ray source observed above few hundreds of GeV is the blazar Mrk\:421, at $z = 0.031$ or $d = 140\:$Mpc, which was detected in 1992 \cite{1992Natur.358..477P}. A single other gamma-ray source had been discovered by the Whipple telescope three years earlier at such energies, the Crab Nebula located within the Milky Way at a distance of about 2\:kpc from Earth \citep{1989ApJ...342..379W}. With a flux in March-June 1992 estimated to 30\% that of the Crab, even a relatively low emission state\footnote{See Ref.~\cite{2010A&A...524A..48T} for historical gamma-ray lightcurves of Mrk\:421.} of Mrk\:421 proved  blazars to be extremely bright gamma-ray beacons distributed on extragalactic scales. Seven other blazars and a single radio galaxy, M\:87, had been detected by 2005 with the Whipple telescope and its successors such as HEGRA (La Palma, Canary Islands) and CAT (Pyrenees, France) \cite{2008ICRC....3.1341W}.\footnote{See \url{http://tevcat2.uchicago.edu/} and references catalogued therein.} 

The late 2000s saw the emergence of a new generation of gamma-ray telescopes, H.E.S.S. (Khomas Highlands, Namibia), MAGIC (La Palma, Canary Islands) and VERITAS (Arizona, USA), with lower energy thresholds down to 100\:GeV, as well as improved background rejection and imaging capabilities (68\% containment angle better than 5 arcmin above 1\:TeV). They revealed a population of TeV extragalactic sources dominated by blazars of the BL\:Lac type (BLL). Unbeamed and steady TeV emission was also discovered from a few galaxies: the radio galaxy Centaurus\:A and the starburst galaxies NGC\:253 and M\:82, all of which are located in the Council of Giants that surrounds the Milky Way at $d=3-6\:\text{Mpc}$ \cite{2014MNRAS.440..405M}. The launch of the \textit{Fermi}-LAT satellite in 2008, with a gain of about a factor of five in effective area and in field of view with respect to EGRET, triggered a tremendous growth in the number of extragalactic gamma-ray sources detected above 100\:MeV. With an angular resolution (68\% containment angle) better than a degree above 1\:GeV \cite{2015CRPhy..16..600T}, cross-identification with radio, optical and X-ray catalogs established the prominence of blazars in the gamma-ray sky, although with marked differences with the emerging population uncovered by ground-based telescopes at higher energies.  
About half of the blazars categorized in the first \textit{Fermi}-LAT catalog (\cite{2010ApJS..188..405A}, 1FGL in Figure~\ref{fig:kifune_xgal}) were found to be flat-spectrum radio quasars (FSRQs), at variance with the prevalence of BLLs at higher energies. The relatively high accretion rate and bright photon fields in the environment of FSRQs with respect to BLLs is nowadays understood as the probable cause for the fainter but higher-energy peak luminosity of BLLs \cite{2017MNRAS.469..255G}.

As shown in Figure~\ref{fig:kifune_xgal}, the number of extragalactic sources identified above 100\:MeV and above 100\:GeV increased by nearly order-of-magnitude in the 2000s. After an initial phase with a high discovery rate, a saturation in number of discovered sources was prevented in the 2010s thanks to improved observing strategies and analysis techniques, continued operation of \textit{Fermi}-LAT and upgrades of H.E.S.S., MAGIC and VERITAS to lower the energy thresholds of the observatories below 100\:GeV. Among the many discoveries enabled by relentless efforts from the gamma-ray community, one could highlight the discovery of the gravitationally-lensed FSRQ B2\:0218+35 located at $z=0.954$ ($d \sim 6000\:$Mpc), a distance record for AGNs observed above 100\:GeV, only recently superseded  by the detection of an FSRQ announced at $z=0.991$ \cite{atel15020}. For comparison, blazars observed by \textit{Fermi}-LAT have been firmly detected above 100\:MeV from beyond $z=3$ and tentative detections have even been reported out to $z>4$ \cite{2020ApJ...903..128K}. 
Other interesting discoveries lie in the few dozen objects above 100\:MeV that do not belong to the classical types of extragalactic gamma-ray sources, namely blazars and radio galaxies. For example, narrow-line Seyfert 1 (NLSY1) galaxies can display jets with an inferred power similar to that of BLLs but with presumably brighter photon fields, as expected in FSRQs \cite{2019Galax...7...87D,2020Univ....6..136F}. The inferred black-hole masses within NLSY1 are estimated around ten million solar masses, rather than billions for numerous blazars and radio galaxies, which would suggest a high accretion rate for NLSY1. Together with compact steep spectrum sources and steep-spectrum radio quasars, NLSY1 could provide one of the missing links in the understanding of jet formation around super massive black holes.

Last but not least, the 2010s and 2020s have seen the emergence of gamma-ray bursts (GRBs) as an entirely new extragalactic population at very-high energies. While the emission of other extragalactic sources is often thought to stem from galactic-scale outflows, be they jets for active galaxies or collective winds for starforming ones, the power engine of GRBs is of stellar size and their outbursts are understood as being induced by the collapse of short-lived massive stars or by a binary merger. Short GRBs, with a duration below a couple of seconds, are subdominant in the gamma-ray band (less than 10\% above 100\:MeV, none above 100\:GeV) with respect to long GRBs \cite{2019ApJ...878...52A}. The latter are now detected up to $z=4.35$ (GRB\:080916C \cite{2009Sci...323.1688A}) by gamma-ray satellites above 100\:MeV and up to $z=1.1$ (tentative redshift of GRB\:201216C \cite{2020GCN.29075....1B}) by gamma-ray telescopes above 100\:GeV.

Multi-wavelength observations of extragalactic gamma-ray sources provide remarkable insights into the non-thermal processes at play in astrophysical environments, as discussed in other reviews of this Special Issue on Extragalactic TeV Astronomy. Of particular interest to the present discussion are source redshifts, whose spectroscopic measurements can be challenging when the emission lines from the host galaxy are overwhelmed by non-thermal emission from the outflows. Despite these difficulties, dedicated spectroscopic campaigns, preceding, coordinated with, or following up the gamma-ray discoveries (see Ref.~\cite{2021A&A...650A.106G} and references therein), have enabled to constrain the distances of over 80\% of the extragalactic gamma-ray sources detected above 100\:GeV. These redshift measurements and gamma-ray detections act as the starting blocks of the journey of gamma-rays over cosmological distances, whose multiple paths are discussed in the following sections. 
 
\section{A gamma-ray journey through cosmic ages}
\label{sec:journey}

\begin{figure}[p]
    \includegraphics[width=1.27\linewidth,right]{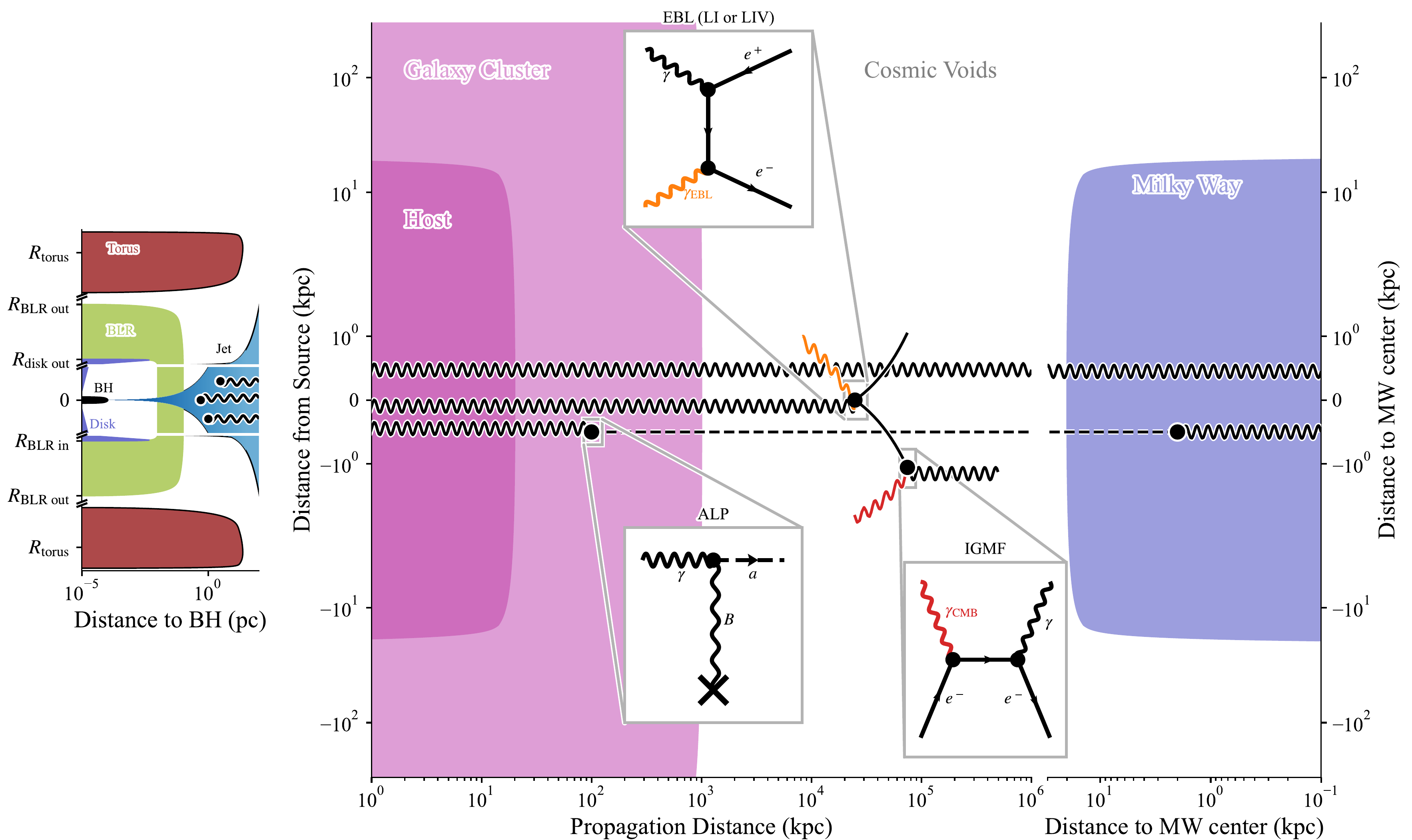}
    \caption{Illustration of gamma-ray propagation from a blazar at luminosity distance $d =  1\:\mathrm{Gpc} $ ($z\approx0.2$) to the Milky Way.
    \textit{Left}: The inner part of a blazar (here an FSRQ) with a black hole of mass $M_\bullet = 10^{9}\:M_\odot$ and a radius equal to the Schwarzschild radius $2\:r_g$ with $r_g = G M_\bullet / c^2 \approx 1.5\times10^{14}\:\mathrm{cm}\approx 4.8\times10^{-5}\:\mathrm{pc}$. The inner and outer radii of the accretion disk, broad line region (BLR), and dusty torus are based on typical values from the literature with 
    $R_\mathrm{disk,~in} = 6\:r_g$, $R_\mathrm{disk,~out} = 200\:r_g$, $R_\mathrm{BLR,~in} = 0.01\:\mathrm{pc}$, $R_\mathrm{BLR,~out} = 0.1\:\mathrm{pc}$, $R_\mathrm{torus,~in} = 5\:\mathrm{pc}$, $R_\mathrm{torus,~out} = 23\:\mathrm{pc}$ (e.g., Ref.~\cite{2016ApJ...830...94F}).
    The jet is assumed to have a parabolic base with a transition into a conical shape at $2\times10^5\:r_g$ (see Eq.~(13) and Figure~1 in Ref.~\cite{2015MNRAS.453.4070P}). 
    Not shown here is the possibility of internal absorption of gamma-rays on the photon fields within the source (e.g., on photons originating from the BLR or the dusty torus). 
    Several observations of FSRQs indicate that gamma-rays are produced outside the BLR~(see, e.g., Refs.~\cite[][]{2019ARA&A..57..467B,2020Galax...8...72C} and references therein). 
    \textit{Center}: Once produced, the gamma-rays propagate through the host galaxy ($r_\mathrm{host} = 20\:$kpc \cite{2016MNRAS.460.3202O}) and potentially through a Galaxy cluster (with $r_\mathrm{cluster} = 1\:\mathrm{Mpc}$ here) before entering cosmic voids. 
    The upper gamma-ray propagates all the way towards the observer whereas the central gamma-ray interacts with an EBL photon to form an electron-positron pair at a distance equal to the mean free path, $\Gamma_{\gamma\gamma} \approx 250\:\mathrm{Mpc}$, of a 1\:TeV gamma-ray. The electron and positron gyrate in the IGMF with $r_\mathrm{gyro} \approx 540\:\mathrm{kpc}\:(\gamma/10^6) (B / 10^{-15}\:\mathrm{G})^{-1}$ and inverse-Compton scatter off CMB photons after a mean free path of $\Gamma_{e\gamma} \approx 730\:\mathrm{kpc}\:(\gamma / 10^6)^{-1}$.
    The lower gamma-ray converts into an ALP in the magnetic field of the galaxy cluster. 
    \textit{Right}: The surviving gamma-ray enters the Milky Way and the ALP converts back to a gamma-ray in the magnetic field of our Galaxy. 
    }
    \label{fig:propagation}
\end{figure}

As described in the introduction, gamma-rays with energies of at least 100\:GeV have now been observed from sources (mostly blazars) over a large range of distances, out to $z \approx 1$ from the ground and $z \approx 4$ from space-borne instruments. The observation of gamma-rays from such distant emitters provides us with a unique opportunity to study the propagation of the highest-energy electromagnetic radiation over cosmological distances \cite{2021Univ....7..146F}. 
In this section, we review the interactions that gamma-rays can undergo along their journey to Earth. 
Just as the absorption of optical emission can be used to infer properties of, e.g., stellar or planetary atmospheres~\cite{Burrows12601}, the absorption of gamma-rays in the intergalactic medium is a complementary tool to study the cumulative energy release by stars and dust grains over the history of the Universe. 
Furthermore, the high energies and long distance scales involved might enable us to observe effects of physics beyond the standard model, such as oscillations into hypothetical scalar particles or the departure from Lorentz invariance. 
The processes discussed in this section are shown in Figure~\ref{fig:propagation}, which illustrates the path of a gamma-ray from its creation in the jet of a blazar (we take here the example of an FSRQ) until it enters the Milky Way. 
For illustration, we follow the path of three gamma-rays produced in the blazar jet (left-most panel). The top gamma-ray travels unperturbed until it is detected on Earth. 
The second gamma-ray in the center is absorbed in intergalactic space, whereas the bottom gamma-ray gets transformed into a hypothetical axion-like particle (ALP) and converts back into a gamma-ray. 
We start our discussion with the absorbed gamma-ray.

It was realized in the 1960s that gamma-rays produced in distant sources should undergo absorption in collisions with photons from background radiation fields~\cite{1967PhRv..155.1408G} such as intergalactic starlight~\cite{Nikishov1962}, radio-frequency fields~\cite{Goldreich1964}, the cosmic microwave background (CMB)~\cite{1966PhRvL..16..479J, 1966PhRvL..16..252G}, or radiation fields within the source~\cite{1966Natur.211..472J}. 
The interaction of a high-energy gamma-ray, $\gamma$, with a photon from a background radiation field, $\gamma_\mathrm{bkg}$, produces an electron-positron pair, $\gamma + \gamma_\mathrm{bkg} \to e^+ + e^-$ if the following threshold condition is met~\cite{1967PhRv..155.1404G}
\begin{equation}
    E'\epsilon' \geq \frac{2\left(m_{\rm e}c^2\right)^2}{1-\cos\theta'}.\label{eq:thr}
\end{equation}
In Eq.~\eqref{eq:thr}, $m_{\rm e}$ is the mass of the electron (and positron), $E'$ and $\epsilon'$  are the energies  of the gamma-ray and background photon, respectively, which are denoted with a prime in the comoving  cosmological frame, and $\theta'$ is the angle between the momenta of the two photons.
The Breit-Wheeler cross section for pair production, $\sigma_{\gamma\gamma}$, is a function of the velocity of the electron (positron) in the center of mass frame with $\beta'^2= 1 - 2(m_e c^2)^2 / [E'\epsilon'(1-\cos\theta')]$ and peaks for $\beta' \approx 0.7$. 
Averaging over the $(1-\cos\theta')$ term, one finds that the peak of the cross section is reached when $\epsilon' \approx 1.0\:\mathrm{eV}(E / \mathrm{TeV})^{-1}$ or for a wavelength $\lambda' \approx 1.2\:\mu\mathrm{m}(E / \mathrm{TeV})$.
As a consequence, gamma-rays at energies $E>100\:$GeV most likely interact with photons at optical to infrared wavelengths. 
The absorption process is shown in the central panel of Figure~\ref{fig:propagation} for the central gamma-ray line.
For a specific photon density $\mathrm{d}n / \mathrm{d}\epsilon$ (in units of eV$^{-1}$\:cm$^{-3}$), 
the mean-free path, $\Gamma_{\gamma\gamma}$, for pair production occurring at a redshift $z$ is given by 

\begin{equation}
    \Gamma_{\gamma\gamma}^{-1}(E',z) = \int_{0}^{\infty} \mathrm{d} \epsilon' \frac{\mathrm{d} n (\epsilon', z)}{\mathrm{d} \epsilon'} \int_{-1}^{1} \mathrm{d}\cos\theta' \frac{1-\cos\theta'}{2} \sigma_{\gamma\gamma}(\beta') \Theta(\epsilon'-\epsilon'_{\rm th}),
\end{equation}
where $\Theta$ is the Heavyside step function which ensures that the threshold condition in Eq.~\eqref{eq:thr} is met. 
In Section~\ref{sec:probing-ig}, we discuss the EBL, which is the radiation field dominating the absorption of gamma-rays. For reference, the mean free path of a 1\:TeV gamma-ray interacting with EBL photons is $\Gamma_{\gamma\gamma} \approx 250\:\mathrm{Mpc}$ for a source at $d = 1\:$Gpc ($z \approx 0.2$) assuming the EBL model in Ref~\cite{2011MNRAS.410.2556D}. This is the distance at which the pair production occurs in the central panel in Figure~\ref{fig:propagation}.
The gamma-ray absorption, quantified by the optical depth $\tau_{\gamma\gamma}$, is then found as a line-of-sight integral up to the redshift $z_0$ of the source over the mean-free path, 
\begin{equation}
    \label{eq:tau}
    \tau_{\gamma\gamma}(E, z_0) = \int_0^{z_0}\Gamma_{\gamma\gamma}^{-1}(E(1+z),z) \frac{\mathrm{d}\ell(z)}{\mathrm{d}z} \mathrm{d}z,
\end{equation}
where $\mathrm{d}\ell/\mathrm{d}z$ is the Jacobian for the transformation of the line-of-sight integral over distance to redshift, which depends on the adopted cosmology. Throughout this review, we assume a flat $\Lambda$CDM cosmology with reference values for the matter density $\Omega_{\rm M}=0.3$, dark-energy density $\Omega_{\Lambda}=0.7$ and Hubble constant at current epoch $H_0 = 70\:\text{km}\:\text{s}^{-1}\:\text{Mpc}^{-1}$, so that $\mathrm{d}\ell/\mathrm{d}z = c/ \Big(H_0(1+z)\sqrt{\Omega_\text{M}(1+z)^3+\Omega_\Lambda}\Big)$.

The initial flux emitted by the blazar, $F_\mathrm{emitted}$,  is  attenuated and we observe the flux 
$F_\mathrm{obs} = F_\mathrm{emitted} \exp(-\tau_{\gamma\gamma})$.
As discussed in Section~\ref{sec:probing-ig}, we can use the observation of sources at different energies and redshifts to reconstruct $\tau_{\gamma\gamma}$ and thus reconstruct the EBL photon density, $\mathrm{d}n / \mathrm{d}\epsilon$. 

In the pair-creation process, the electrons and positrons acquire about half of the initial gamma-ray energy, so that their Lorentz factors reach $\gamma = E / (2m_e c^2) \approx 10^6\ (E / \mathrm{TeV})$. If we follow the path of the electrons and positrons in Figure~\ref{fig:propagation}, we see that they can undergo an inverse-Compton scattering process with background radiation fields.  The most abundant photon field is the CMB with an integrated energy density of $U_\mathrm{CMB}\approx 0.26\:\mathrm{eV}\:\mathrm{cm}^{-3}$ and an average photon energy of $\langle \epsilon_\mathrm{CMB} \rangle \approx 630\:\mu\mathrm{eV}$. For such CMB photons and for electrons at $\gamma \approx 10^6$, the inverse-Compton scattering occurs in the Thomson regime, i.e.\ the electrons only loose a small fraction of their energy in each scattering event (see, e.g., \cite{1970RvMP...42..237B}). In the Thomson regime, the mean free path $\Gamma_{e\gamma}$ and the average photon energy after scattering, $\langle\tilde{\epsilon}\rangle$, are
\begin{eqnarray}
\label{eq:MFP_egammma}
\Gamma_{e\gamma} &=& ct_\mathrm{cool} = \frac{c\gamma}{|\dot{\gamma}|} = \frac{3m_e c^2}{4\sigma_\mathrm{T} U_\mathrm{CMB} \gamma}\approx 0.7\:\mathrm{Mpc}\ \left(\frac{E}{\mathrm{TeV}}\right)^{-1},\\
\label{eq:Esecond}
\langle\tilde{\epsilon}\rangle &=& \frac{4}{3}\langle\epsilon_\mathrm{CMB}\rangle\gamma^2 \approx 0.8\:\mathrm{GeV}\ \left(\frac{E}{\mathrm{TeV}}\right)^2,
\end{eqnarray}
where $\sigma_\mathrm{T} \approx 6.65\times10^{-25}\:\mathrm{cm}^2$ is the Thomson cross section and $t_\mathrm{cool} = \gamma / |\dot{\gamma}|$ is the cooling time. 
Two observations can be made from these equations. On the one hand, for a sufficiently large primary gamma-ray energy, $E$, the upscattered CMB photon can reach an energy $\langle\tilde{\epsilon}\rangle$ that is in the gamma-ray regime, so that the secondary gamma-ray could again pair produce on a background photon and initiate a full electromagnetic cascade. 
On the other hand, Eq.~\eqref{eq:MFP_egammma} indicates that the electron-positron pairs can travel over hundreds of kpc before they scatter. As they are charged, they are deflected in magnetic fields along the line of sight. 
A sufficiently large magnetic field would then broaden the pair beam. The resulting gamma-ray cascade would arrive with a time delay with respect to the primary gamma-ray and could even form an extended halo around the otherwise point-like blazars~\cite{1994ApJ...423L...5A,1995Natur.374..430P}. These effects have been suggested as potential probes of the strength of the intergalactic magnetic field (IGMF), which is extremely difficult to observe directly (see, e.g., Ref.~\cite{2013A&ARv..21...62D}). The current constraints are discussed in Section~\ref{sec:probing-ig}.

We now turn to processes beyond the standard model that can affect gamma-ray propagation. 
One possibility is that gamma-rays oscillate into axions 
in the presence of external electromagnetic fields. 
The axion was originally proposed to explain the apparent absence of the violation of parity symmetry (P) and time reversal (T) in QCD~\cite{1977PhRvD..16.1791P,1978PhRvL..40..223W,1978PhRvL..40..279W}. It was soon realized that this axion is also a dark-matter candidate~\cite{1983PhLB..120..137D,1983PhRvL..51.1415S,1983PhLB..120..133A,2012JCAP...06..013A} and that particles with similar properties, so-called axion-like particles (ALPs), arise in extensions of the standard model, most notably string theory (see, e.g., Ref.~\cite{2010ARNPS..60..405J} for a review). 
The interaction between axions (or ALPs) and photons arises through the Lagrangian density $\mathcal{L}_{a\gamma} = g_{a\gamma} \mathbf{E}\cdot\mathbf{B} a$,\footnote{To see how axions change Maxwell's equations, we refer the reader to Ref.~\cite{2013MPLA...2850162V}.} where $g_{a\gamma}$ is the coupling between axions and photons (in units of inverse energy), $\mathbf{E}$ is the electric field associated to the photon, $\mathbf{B}$ is an external magnetic field, and $a$ is the axion field strength~\cite{1983PhRvL..51.1415S,1988PhRvD..37.1237R}. 
This Lagrangian term describes the photon-axion Primakoff effect which, similarly to neutrino flavor mixing, results in an oscillation between photons and axions in the presence of external fields. 

If produced through the so-called misalignment mechanism in order to make up a fraction $\Omega_{a,~\mathrm{DM}}$ of the total dark matter density (as determined, e.g., with the \textit{Planck} satellite to be $\Omega^{Planck}_\mathrm{DM} = 0.26$~\cite{2020A&A...641A...6P}), the photon-ALP coupling must fulfill the relation (adopted from Ref.~\cite{2012JCAP...06..013A}),

\begin{equation}
    g_{a\gamma} < 2.5\times10^{-11}\:\mathrm{GeV}^{-1} \theta_1 \mathcal{N} \left(\frac{m_a}{\mu\mathrm{eV}}\right)^{\frac{1}{2}}\left(\frac{\Omega_{a,~\mathrm{DM}}}{\Omega_\mathrm{DM}^{Planck}}\right)^{-\frac{1}{2}},
\end{equation}
where $\theta_1$ is the initial misalignment angle of the ALP field in the early Universe (which takes values between $-\pi$ and $\pi$) and $\mathcal{N}$ is a model dependent parameter; in the simplest case, we take $\theta_1\mathcal{N} \sim 1$.

The Feynman diagram for the photon-axion interaction is shown in the bottom central panel of Figure~\ref{fig:propagation}.
We used the example of a gamma-ray converting into an ALP in the magnetic field of a galaxy cluster that might harbor the blazar. The ALP can then convert back into a gamma-ray in, e.g., the magnetic field of the Milky Way, as discussed in Section~\ref{sec:pbsm}.
Photon-axion oscillations could be detected in gamma-ray observations of blazars mainly in two ways. 
On the one hand, energy-dependent oscillations could be visible around two specific energies $E_\mathrm{crit}^\mathrm{low}$ and $E_\mathrm{crit}^\mathrm{high}$ (see, e.g., Ref.~\cite{2007PhRvL..99w1102H,2010JCAP...05..010B} with the addition from Ref.~\cite{2015PhRvD..91h3003D}),
\begin{eqnarray}
E_\mathrm{crit}^\mathrm{low} &=& \frac{|m_a^2 - \omega_\mathrm{plasma}^2|}{2g_{a\gamma} B}\nonumber\\
&\approx& 2.6\:\mathrm{GeV}\ \left(\frac{g_{a\gamma}}{10^{-11}\:\mathrm{GeV}^{-1}}\right)^{-1}
\left(\frac{B}{\mu\mathrm{G}}\right)^{-1}\left|\left(\frac{m_a}{\mathrm{neV}}\right)^2 - 1.4\times10^{-4}\left(\frac{n_\mathrm{el}}{\mathrm{cm}^{-3}}\right)\right|\label{eq:ecrit-low}\\
E_\mathrm{crit}^\mathrm{high} &=& \frac{g_{a\gamma}B}{\chi_B+\chi_\mathrm{photon}}\nonumber\\
&\approx&2.1\times10^{6}\:\mathrm{GeV}\ \left(\frac{g_{a\gamma}}{10^{-11}\:\mathrm{GeV}^{-1}}\right)\left(\frac{B}{\mu\mathrm{G}}\right)\left[\left(\frac{B}{\mu\mathrm{G}}\right)^2 + 5.5\left(\frac{\chi_\mathrm{photon}}{\chi_\mathrm{CMB}}\right)\right]^{-1},\label{eq:ecrit-high}
\end{eqnarray}
where $m_a$ is the mass of the axion; $\omega_\mathrm{plasma}$ is the plasma frequency of the traversed medium with electron density $n_\mathrm{el}$; $\chi_B = \alpha/(45\pi)(B/B_\mathrm{cr})^2$ where $B_\mathrm{cr} \approx 4.4\times10^{13}\:\mathrm{G}$ is the critical magnetic field and $\alpha$ is the fine-structure constant; $\chi_\mathrm{photon}$ describes the strength of the dispersion of the photon from background radiation fields. 
If the only relevant field is the CMB with $T_\mathrm{CMB} \approx 2.73\:\mathrm{K}$, then $\chi_\mathrm{photon} = \chi_\mathrm{CMB} = (k_B T_\mathrm{CMB})^4 (44\pi^2/2025) \alpha^2/m_e^{4} \approx 5.1\times10^{-43}$~\cite{2015PhRvD..91h3003D}.
Below $E_\mathrm{crit}^\mathrm{low}$, the oscillations are suppressed due to a momentum mismatch between the (massive) axion and the photon with an effective mass equal to the plasma frequency. 
Above $E_\mathrm{crit}^\mathrm{high}$, the mixing is suppressed as well: photon-photon dispersion (i.e.\ higher order corrections to the photon propagator) with the magnetic field and / or background radiation fields becomes more likely than photon-ALP oscillations. 
On the other hand, for gamma-ray energies $E_\mathrm{crit}^\mathrm{low} < E < E_\mathrm{crit}^\mathrm{high}$ the mixing becomes maximal and independent of energy. 
For a path length $L$ in a homogeneous magnetic field, the oscillation probability is simply given by $P_{a\gamma} \approx (2\pi L / \lambda_\mathrm{osc})^2$ with the oscillation length $\lambda_\mathrm{osc} \sim 4\pi / (g_{a\gamma} B) \sim 0.4\:\mathrm{Mpc}\ (B / \mu\mathrm{G})^{-1} (g_{a\gamma} / 10^{-11}\:\mathrm{GeV}^{-1})^{-1}$.
Interestingly, in this strong mixing regime, a fraction of the gamma-rays could avoid pair production by converting into axions if $\lambda_\mathrm{osc} \lesssim \Gamma_{\gamma\gamma}$. If these axions re-convert into photons close to Earth, the observed photon flux would be enhanced in comparison to the no-axion case and the Universe would appear more transparent to high-energy gamma-rays than expected from conventional physics only.  
The status for axion searches using both effects is reviewed in Section~\ref{sec:pbsm}. 

Lastly, gamma-ray propagation could also be altered if Lorentz invariance is violated. 
Such a Lorentz invariance violation (LIV) is expected from theories that try to unify quantum field theory with gravity~(see Ref.~\cite{2021arXiv211105659A} and references therein). 
A recent review of the astrophysical effects of LIV is given in Refs.~\cite{2020Symm...12.1232M,2021Univ....7..345T}, which we loosely follow hereafter.
One of the consequences of LIV would be that the photon velocity is not always the speed of light, $c$, but depends on the energy of the photon. 
Some effective-field theories predict that LIV can lead to a modified dispersion relation for particles at leading order $n$~\cite{1998Natur.393..763A}. 
For photons with energy $E$ and momentum $p$, this can be written as 

\begin{equation}
\label{eq:MDR}
E^2 - p^2 = \pm (|\xi_n| / M^n) E^{n+2},
\end{equation}
where $\xi_n$ is a coefficient of the underlying theory and $M$ is the energy scale at which new physics is expected, e.g., the Planck energy scale $E_\mathrm{Pl} = \sqrt{\hbar c^5 / G} \approx 1.22\times10^{19}\:\mathrm{GeV}$. Tight bounds have been placed on the superluminal case (plus sign in Eq.~\eqref{eq:MDR}), in particular by recent observations by HAWC and LHAASO of gamma-rays beyond 100\:TeV from sources in the Milky Way \cite{2020PhRvL.124m1101A, 2021arXiv210612350T}. We focus the discussion here on subluminal searches, i.e.\ a negative sign in Eq.~\eqref{eq:MDR}, which correspond to velocities below the speed of light for photons. Within specific models predicting LIV, such as the Standard-Model Extension \cite{1997PhRvD..55.6760C}, corrective terms with an even order $n$ emerge from operators that are invariant under charge/parity/time-reversal (CPT) symmetries, contrarily to terms with an odd order $n$ \cite{1998PhRvD..58k6002C,2009PhRvD..80a5020K}. Quadratic modifications of the dispersion relation may thus be favored over linear modifications within CPT-invariant theories.

In terms of gamma-ray propagation, the modified dispersion relation has two important consequences. 
On the one hand, the threshold for pair production is modified by adding the additional term $|\xi_n| E'^{n+2} /(4M^n)$ to the right hand side of Eq.~\eqref{eq:thr},
where $\xi_n = 1$ if LIV only affects photons and $\xi_n = (1 - 2^n)^{-1}$ if it affects both photons and electrons~\cite{2009NJPh...11h5003S,2009ApJ...691L..91S}.
As a result, pair production will cease once the ratio $E'^{n+2} / M^n$ becomes large compared to $(2m_e c^2)^2$. The latter two terms are comparable at $E' \sim 20\:\mathrm{TeV}$ for $M = E_\mathrm{Pl}$ and $n=1$, so that the optical depth $\tau_{\gamma\gamma}$ would decrease towards zero at gamma-ray energies larger than $E'$. Thus, if one can measure blazars beyond these energies, it should be possible to probe LIV.

On the other hand, an energy-dependent speed of light would lead to a time delay $\Delta t$ between two photons emitted simultaneously with energy difference $\Delta E > 0$ propagating from a source at redshift $z$ (e.g., Ref.~\cite{2008JCAP...01..031J} but see also Ref.~\cite{2015PhRvD..92l4042R} for an alternative formulation),

\begin{equation}
    \Delta t = \frac{1+n}{2H_0}\frac{\xi_n^n \Delta E^n}{M^n} \int_0^z \frac{(1 + z')\mathrm{d}z'}{\sqrt{\Omega_\text{M} (1 + z')^3 + \Omega_\Lambda}} \approx
    \begin{cases}
    18\:\mathrm{s} \left(\frac{\Delta E}{\mathrm{TeV}}\right) \left(\frac{M}{\xi_n E_\mathrm{Pl}}\right)^{-1} F_{1}(z) & n=1, \\
    66\:\mathrm{s} \left(\frac{\Delta E}{\mathrm{TeV}}\right)^2 \left(\frac{M}{\xi_n 10^{20}\:\mathrm{eV}}\right)^{-2} F_{2}(z) & n=2.
    \end{cases}
\end{equation}
We have approximated the integral with polynomials $F_n(z) = \sum_{k = 1}^{n+1} p_k z^k$ of the $(n+1)$-th order for $0.01 \leqslant z \leqslant 4$ which show an absolute error of less then 4\:\% compared to the numerical integration. In the case $n = 1$, $F_1(z)$ is almost linear with $F_1(z)\approx 1.2\: z$ whereas, for $n=2$, $F_2(z)$ increases from $\approx 10^{-4}$ for $z=0.01$ to $\approx 9$ for $z=4$.\footnote{
For $n=1$, the coefficients are $p_2 = -0.05,~p_1 = 1.22,~p_0 = -0.04$ and for $n=2$ one finds 
$p_3 = -0.04,~p_2 = 0.58,~p_1= 1.20,~p_0=-0.04$.
}
Sources as distant as possible and electromagnetic radiation over a wide (linear) energy range are preferred to maximize the potentially observable effect. As discussed in Section~\ref{sec:pbsm}, gamma-ray bursts and blazar flares are best suited for these kinds of searches. 

\section{Probing the content of the intergalactic medium}
\label{sec:probing-ig}

\begin{figure}[ht!]
    \includegraphics[width=1.11\linewidth,right]{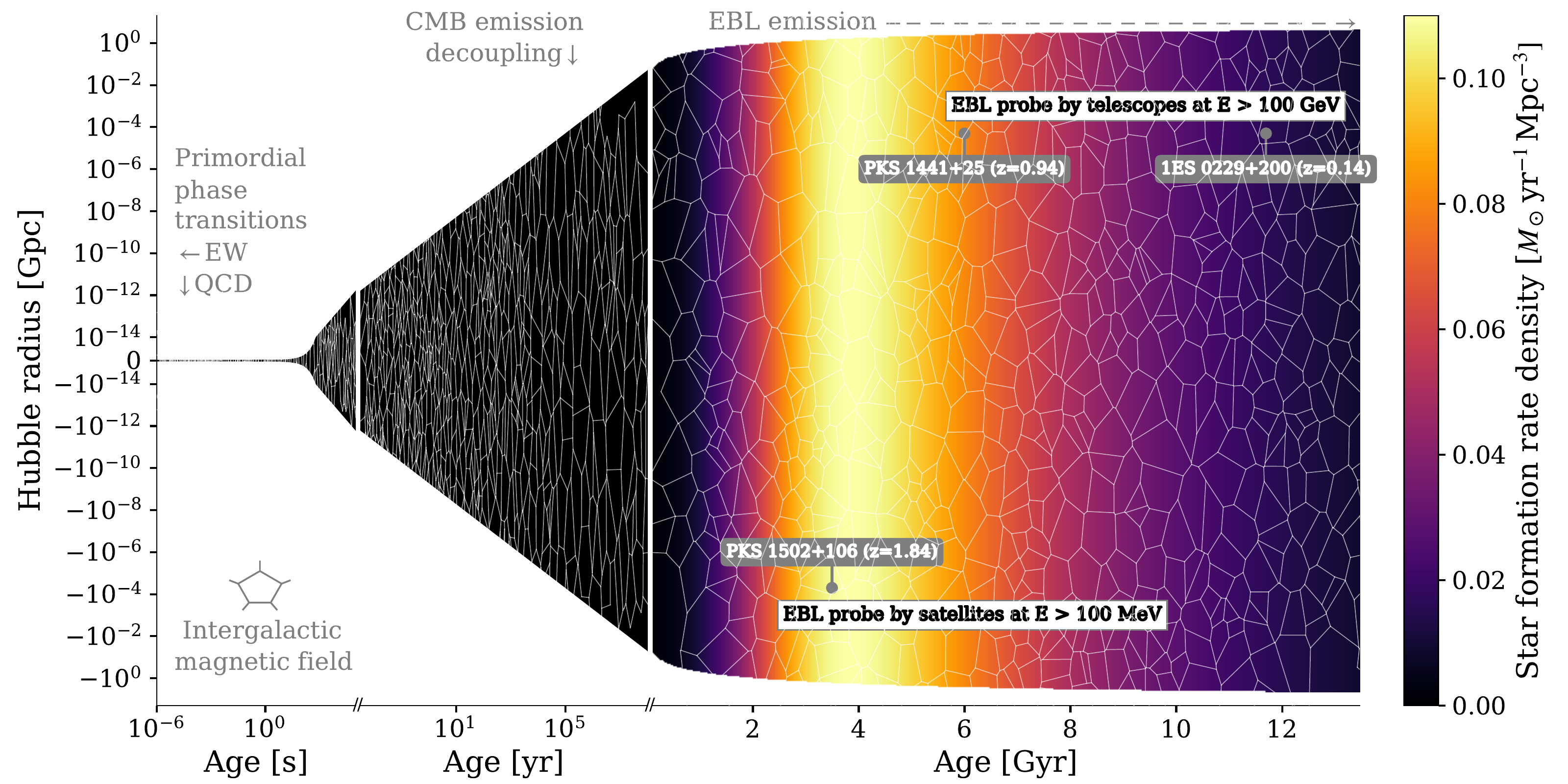}
    \caption{Evolution within a Hubble radius of the two classical entities relevant to gamma-ray propagation, the IGMF and the EBL, as a function of cosmic age. The redshift range over which constraints on gamma-ray propagation have been established by ground-based telescopes ($E>100\:$GeV) is illustrated by the upper white box in the right-hand-side panel. The latter observations are exemplified by the extreme blazar of BLL type 1ES\:0229+200 ($z=0.14$), which has been seminal for IGMF studies \citep{2010Sci...328...73N, 2011A&A...529A.144T}, and by the most distant TeV blazar used to constrain the EBL, the FSRQ PKS\:1441+25 \citep{2015ApJ...815L..22A,2015ApJ...815L..23A}. The redshift range over which constraints have been established by satellite-based observations  ($E>100\:$MeV) is illustrated by the lower white box, exemplified by the FSRQ-type PKS\:1502+106 ($z=1.84$) which is the most-distant blazar detected by \textit{Fermi}-LAT above 30\:GeV \citep{2020ApJ...892..105A} with a firm spectroscopic redshift \citep{goldoni_paolo_2021_5512660}. The emission of the EBL from a cosmic age of about 1\:Gyr is illustrated by the colored area, which figures the star-formation rate density inferred from gamma-ray observations \cite{2018Sci...362.1031F}, as parameterized by Ref.~\cite{2021MNRAS.503.2033K}. Earlier phases of the expansion of the Universe are illustrated in the middle panel, which features the decoupling that resulted in the CMB emission, and in the left-hand-side panel, which includes the quantum chromodynamics (QCD) phase transition (or cross-over) \citep{1989ApJ...344L..49Q} while the electroweak (EW) phase transition (or cross-over) \cite{2021RPPh...84g4901V} occurred at earlier times. As illustrated by the magnetic cells shown in the first two panels, both the EW and QCD eras could have seen the IGMF emergence, but it should be noted that the IGMF could also have been seeded at later epochs during the formation of large-scale structures. The reader is referred to Figure~3 in Ref.~\cite{PueschelBiteau22} and Figure~9 in Ref.~\cite{2021Univ....7..223A} for more specific constraints on the EBL spectrum and IGMF parameter space, respectively.}
    \label{fig:ebl_igmf_evol}
\end{figure}

The baryonic and electromagnetic contents of the Universe can be probed through observations of extragalactic gamma-ray sources at various distances, as illustrated in Figure~\ref{fig:ebl_igmf_evol}. The first entity of interest consists of all ultra-violet to infrared photons emitted since reionization, from a time when the Universe was about 1\:Gyr old to the current epoch, about 13\:Gyr later. Such photons accumulate in the intergalactic medium to build up the main constituent of astrophysical origin in the ``spectrum'' of the Universe \citep{2018ApSpe..72..663H}: the EBL. The EBL energy density, ${\rm d} n / {\rm d} \epsilon$, is composed of two humps each spread over nearly two decades in energy: the cosmic infrared background (CIB) at wavelengths $8\:\mu\text{m}< \lambda <1000\:\text{mm}$ or energies $0.16\:\text{eV} >\epsilon > 1.2\:\text{meV}$, which mainly arises from emission by dust grains, and the cosmic optical background (COB) at wavelengths $0.09\:\mu\text{m}< \lambda <8\:\mu\text{m}$ or energies $13.6\:\text{eV} >\epsilon > 0.16\:\text{eV}$, which mainly stems from stellar radiation that escaped its environment without absorption on dust grains. As such, the evolution of the EBL with redshift provides an integrated measure of the evolving star-formation rate density, also called the cosmic star-formation history (CSFH \citep{2014ARA&A..52..415M}).  The second entity of interest is the magnetic field which pervades the space between filaments and clusters of galaxies: the IGMF \citep{2013A&ARv..21...62D}. The IGMF is a crucial ingredient to understand cosmic magnetism on both large and small spatial scales, particularly as it could have provided the seed later amplified through compression and dynamo effects into the $\mu$G fields observed in galaxies and clusters \citep{2010A&A...513A..30B}. The IGMF could originate both from astrophysical outflows (AGN jets, starburst winds) or primordial phase transitions, e.g., at the break of the electroweak symmetry \cite{2021RPPh...84g4901V}. The reader is referred to Ref.~\cite{PueschelBiteau22} for a didactic review on these matters.

As discussed in Section~\ref{sec:journey}, gamma-rays can interact with photons from the EBL en route to the observer, which results in an absorption of the emitted spectrum. The magnitude of the absorption depends both on the cross section of interaction and on the density of target EBL photons in the wavelength range of interest. For example, spectral observations of nearby blazars such as Mrk\:421 and Mrk\:501 ($z \sim 0.03$) at 20\:TeV probe the EBL density at typical wavelengths ranging in 10--100\:$\mu$m, that is a fraction of the CIB, while observations at 100\:GeV of sources at $z=1$ typically probe the COB density at wavelengths ranging in 0.1--1\:$\mu$m. Gamma-ray observations of sufficiently bright sources at different redshifts thus enable the study of the EBL intensity over nearly three decades in energy (0.1--100\:$\mu$m). The main limitation of the indirect gamma-ray measurement technique lies in the unknowns on the emitted spectrum, which were historically solved as follows.

The first studies of the EBL based on observed gamma-ray spectra were developed after the discovery of Mrk\:421 by the Whipple observatory. The TeV spectrum was jointly modeled with the unabsorbed spectral measurement from EGRET at MeV-GeV energies, assuming that the emitted spectrum follows a straight power-law dependence with energy from MeV to TeV energies \cite{1994Natur.369..294D}. As soon noted by Ref.~\cite{1994ApJ...436..696D}, such an approach neglects any possible intrinsic curvature in the emitted spectrum, which could be attributed either to absorption inside the source itself rather than on the line of sight or to the interplay between acceleration and radiative processes that would reduce the flux at very-high energies. Although only weak upper limits on the EBL density could be placed with the first observations, constraining limits were derived a decade later with the observations of the extreme blazars H\:2356+304 ($z=0.165$) and 1ES\:1101-232 ($z=0.186$) by H.E.S.S. \citep{2006Natur.440.1018A}. At the lack of MeV-GeV observations to constrain the maximum hardness of the emitted spectrum, the authors of Ref.~\citep{2006Natur.440.1018A} used a theoretical bound on the index of the power-law spectrum and managed to solve a long-standing discrepancy between direct observations of dark patches of the night sky and the integrated galaxy light (IGL), which is determined from counting galaxies in deep-field surveys. Direct observations encompass not only the EBL but also one-to-two order-of-magnitude brighter foregrounds (zodiacal light from dust grains in the Solar system, stellar light from our Galaxy), so that a percent bias in foreground subtraction could result in a 100\% overestimation of the EBL \citep{2019ConPh..60...23M}. Galaxy counts, on the other hand, rely on well-calibrated but limited samples of galaxies. These surveys could fail to account for low-surface-brightness galaxies below the sensitivity threshold of the instrument or miss any diffuse component of the EBL \citep{2021arXiv210212089D}. Up to the late 2000s, the most optimistic direct measurements exceeded the IGL by over an order of magnitude. This ``optical controversy'' \citep{2021arXiv210212089D} appeared to be solved by indirect constraints from H.E.S.S., with gamma-ray upper limits suggesting that most of the COB was resolved in known galaxies.

The ``optical controversy'', or more specifically the question of the actual intensity of the COB between $0.5\:\mu$m and $1\:\mu$m, is not simply a matter of discrepant measurement techniques (see Refs.~ \cite{2021arXiv210212089D,2021ApJ...906...77L} for a recent status of the scientific debate). This region of the EBL spectrum could contain contributions from the sources of reionization (UV photons at ${\sim}\:0.1\:$eV emitted at $z\sim10$), from gas or stars stripped from their host galaxies, e.g., during mergers, or radiative signatures from exotic particles such as decaying or interacting dark-matter candidates. Moreover, the limitations of the gamma-ray indirect technique, namely the unknowns in the emitted spectrum, left trying questions open. Is the theoretical bound on intrinsic hardness sufficiently motivated? With only gamma-ray upper limits at hand, would an inferred EBL level below the IGL suggest a too large gamma-ray transparency, as expected from exotic processes? 

The conceptual ditch was crossed after the launch of \textit{Fermi}-LAT in 2008. Broad-band measurements above $100\:$MeV revealed that the unabsorbed gamma-ray spectra of blazars do show intrinsic curvature and that it can be disentangled from line-of-sight processes \citep{2012Sci...338.1190A}. The techniques employed to account for intrinsic curvature in the first gamma-ray measurements of the EBL \citep{2012Sci...338.1190A,2013A&A...550A...4H} slightly differed, as discussed in Ref.~\cite{2019A&A...627A.110B}. In Ref.~\citep{2012Sci...338.1190A}, the \textit{Fermi}-LAT Collaboration measured intrinsic curvature through a successful modeling of the unabsorbed part of each spectrum in their sample with a log-parabola function (parabola in log-log space, when a power-law model is a linear function in log-log space). The authors then modeled the absorbed part as $F_\mathrm{obs} = F_\mathrm{emitted} \: \exp[-a \tau_{\text{ref}}(E, z_0)]$, by fixing the parameters of the emitted spectrum based on the lower-energy observations and letting free the normalization, $a$, of reference EBL models. In Ref.~\citep{2013A&A...550A...4H}, the H.E.S.S. Collaboration, having a more limited access to the emitted spectrum, left both the EBL normalization and intrinsic parameters free to vary. The optimal level of curvature was determined by testing several curved variations of power-law models and by adopting the preferred one in a frequentist approach (maximization of a goodness-of-fit estimator). The absorption feature, which shows a characteristic dependence on energy and redshift, could thus be reconstructed by combining observations from multiple sources at different redshifts, with a $6\sigma$ signal driven by blazars at $z \approx 0.5-1.6$ observed by \textit{Fermi}-LAT up to ${\sim}\:500\:$GeV and a $9\sigma$ signal driven by blazars at $z \approx 0.1-0.2$ observed by H.E.S.S. up to ${\sim}\:5\:$TeV. The specific intensity of the COB, $\nu I_\nu = \frac{c}{4\pi}\: \epsilon^2 {\rm d} n / {\rm d} \epsilon$ (units of nW\:m$^{-2}$\:sr$^{-1}$), was determined to be at a level compatible with the IGL within statistical and systematic uncertainties, on the order of $20-30\%$ each. H.E.S.S. and \textit{Fermi}-LAT confirmed, for the first time through a measurement, previous exclusions of the most optimistic direct measurements \cite{2006Natur.440.1018A, 2012A&A...542A..59M}. The analysis techniques have been refined over the past decade and larger datasets have been scrutinized. Ref.~\citep{2015ApJ...812...60B} compiled a database of 86 archival spectra from 30 sources observed by all ground-based instruments, together with satellite-based constraints on the emitted spectrum, and detected the absorption signature at the $11\sigma$ level based on an EBL scaling approach similar to those employed by H.E.S.S. and \textit{Fermi}-LAT. A model-independent approach was also developed in Ref.~\citep{2015ApJ...812...60B} to measure the EBL intensity in four distinct wavelength bands spanning $0.3-100\:\mu$m. Imposing that the inferred EBL spectrum does not drop below the IGL level provided a refined measurements with 8 bins over the same wavelength range. Both EBL-model-dependent and -independent techniques were subsequently employed by the H.E.S.S. \citep{2017A&A...606A..59H},  VERITAS \citep{2019ApJ...885..150A} and MAGIC \citep{2019MNRAS.486.4233A} Collaborations, the latter also including \textit{Fermi}-LAT constraints on the intrinsic spectrum. Although the overall EBL intensity can now be estimated from ground-based gamma-ray observations with a statistical accuracy of 10-20\% and a systematic one of 20-30\%, wavelength-resolved measurements come with a poorer resolution, particularly in the region of the ``optical controversy''. The status of the controversy at the time of writing is the following. The New Horizons mission recently performed a direct measurement of the COB beyond Pluto's orbit \cite{2021ApJ...906...77L}, estimating the EBL intensity at 0.6$\:\mu$m to $15.9\pm4.2\:$nW\:m$^{-2}$\:sr$^{-1}$ (stat. + sys.). This value can be compared to the most accurate IGL estimate to date at a comparable wavelength (r-band), $8.11\pm0.33\:$nW\:m$^{-2}$\:sr$^{-1}$ (stat. + sys.), with an impressive accuracy on the order of 5\% obtained through the combination of the GAMA and DEVILS deep-field surveys with observations from the Hubble space telescope \citep{2021MNRAS.503.2033K}. The tension between galaxy counts and direct measurements has thus been reduced from a factor of ten in the early 2010s down to a factor of two, with a low-level significance if we loosely treat the uncertainties as random ones. For comparison, the latest wavelength-resolved measurements from H.E.S.S. and VERITAS limit the maximum EBL intensity at comparable wavelengths to $15-25\:$nW\:m$^{-2}$\:sr$^{-1}$ (sys.\ limited), with estimates that, although favoring galaxy counts, do not settle the debate and leave ample room for unresolved contributions.

An alternative approach to the forward-folding gamma-ray spectral analysis described above has been developed over the past years, particularly to explore the evolution of the EBL with redshift and to constrain cosmological parameters. The EBL density indeed results from the accumulated luminosity density or specific emissivity, $j$ (units of cm$^{-3}$\:s$^{-1}$), of each redshift layer since the ignition of the first stars, so that ${\rm d} n / {\rm d} \epsilon = \big((1+z)^3/c\big) \int_z^\infty \mathrm{d} z' \: \mathrm{d} \ell/\mathrm{d} z'\: j(\epsilon',z')/\epsilon'$, where $\epsilon' = (1+z')\epsilon$ \cite{2018Sci...362.1031F}. For a given luminosity density, the mean dust extinction, $A_\epsilon$, and the amount of light emitted per star-formation-rate unit, $K_\epsilon$ (units of $M_\odot$), provides an estimate of the CSFH, $\rho(z) = K_\epsilon\:10^{0.4A_\epsilon}\: j(\epsilon,z)$ (units of $M_\odot$\:yr$^{-1}$\:Mpc$^{-3}$). With an estimate of the optical depth $\tau_{\gamma\gamma}$ (see Eq.~\eqref{eq:tau}), gamma-ray spectral observations could constrain both the evolution of light emission through cosmic ages, $\rho(z)$, and parameters of the $\Lambda$CDM cosmological model, in particular the Hubble constant via the distance element $\mathrm{d} \ell/\mathrm{d} z$ (see Ref.~\cite{PueschelBiteau22} for a review). Given the complexity of the emissivity-based calculation of the optical depth, which involves a double integral over redshift and an integral over EBL photon energies, the authors of Refs.~\cite{2018Sci...362.1031F,2019ApJ...874L...7D} summarized ensembles of gamma-ray spectra (either from satellite- and/or ground-based observations) into estimates of the optical depth in successive energy bins, by fitting a scaling factor with respect to reference EBL models. The \textit{Fermi}-LAT Collaboration in particular obtained the first gamma-ray indirect measurement of the CSFH through this technique \cite{2018Sci...362.1031F}. Their CSFH and EBL estimates show good agreement with galaxy surveys (see also Ref.~\cite{2021MNRAS.503.2033K} for an independent discussion). While the gamma-ray constraints mostly come from blazars at $z<3$, the integral nature of the gamma-ray measurements provides sensitivity to EBL emission at higher redshifts. The \textit{Fermi}-LAT Collaboration estimated a UV luminosity density above $z=4$ in line with the lowest values from Lyman-break galaxy surveys and some room remains at $z\sim6$ for a sufficiently bright UV field to drive reionization. In Ref.~\cite{2019ApJ...885..137D}, the authors also employed estimates of the optical depth to constrain the Hubble constant and reported an accuracy of 3--6\:km\:s$^{-1}$\:Mpc$^{-1}$ on $H_0$, where the bounds correspond either to a fixed or free matter density, $\Omega_\text{M}$. Such accuracy is relevant to test current tensions between late and early-type measurements of $H_0$, typically distant by 5\:km\:s$^{-1}$\:Mpc$^{-1}$ with a tension at a significance level beyond $4\sigma$ \cite{2021CQGra..38o3001D}. The latest gamma-ray measurements are enticing as they open new scientific avenues to gamma-ray cosmology. Nonetheless, it can be noted that direct estimates of the optical depth thus far rely on scaled versions of EBL models that in turn depend on the CSFH and cosmological parameters, so that estimates of the latter are in principle not free from potential biases in the EBL models themselves (see, e.g., Refs.~\cite{2017A&A...603A..34F, 2018MNRAS.474..898A, 2019MNRAS.484.4174K} for recent models). One of the main question in this line of research during the upcoming decade, independently from the adopted method, lies in the control on systematic uncertainties, which will require improvement upon the typical 20\% resolution on the overall EBL intensity achieved as of today.

In a nutshell, the study of the EBL absorption signature in extragalactic gamma-ray spectra has shifted, over the past decade, from a detection quest to the pursuit of precision measurements. The observational proof of gamma-ray interactions with the EBL naturally leads to the question of how the produced electron-positron pairs dissipate their energy and of the potential observational signatures of such an energy release. As discussed in Section~\ref{sec:journey}, a prime candidate process for dissipation is inverse-Compton scattering with background photon fields. With a bolometric intensity about twenty times larger than that of the EBL, the CMB provides copious amounts of target photons that can be upscattered to GeV energies by the secondary charged particles (see Eq.~\eqref{eq:Esecond}). As shown in Eq.~\eqref{eq:MFP_egammma}, such charged secondaries are produced at distances tens to hundreds of times larger than the typical Mpc scale that corresponds to the radius of the cluster or thickness of the filament hosting the gamma-ray source, so that it was early noted \cite{1995Natur.374..430P} that the detection of a GeV secondary signal could be used as a probe of the IGMF. The latter denotes the largely unknown magnetic fields in cosmic voids, whose volume filling factor is dominant in the cosmic web (see Ref.~\cite{2009LNP...665..291V} for a didactic review on voids). An alternative candidate process to inverse-Compton scattering lies in the development of plasma instabilities in the beam of electrons and positrons (see dedicated overview in Ref.~\cite{2021Univ....7..223A} and references therein). The nature and growth of these instabilities depend on the properties of the intergalactic medium (temperature, density) and on the intensity, energy distribution and angular distribution of the pairs in the beam (see Ref.~\cite{2018ApJ...857...43V} for state-of-the-art simulations and analytical treatment). The present understanding of the competition between the two cooling mechanisms remains limited. If plasma instabilities were to develop faster than inverse-Compton cooling and over a sufficiently long time (hundreds of years based on Eq.~(52) in Ref.~\cite{2018ApJ...857...43V}), the relaxation of the beam could contribute to the heating of the intergalactic medium \cite{2012ApJ...752...23C, 2012ApJ...752...24P,2015ApJ...811...19L}. Observational constraints on the temperature history of the intergalactic medium nonetheless tightly limit the contribution of the secondary pairs, e.g., to the reionization of helium at $z>2-3$ \cite{2016MNRAS.460.1885U}. Clear-cut observational signatures of energy losses through plasma instabilities thus currently remain out of reach.

On the other hand, the dissipation of the energy of the pairs through inverse-Compton cooling would suggest the presence of a bump at GeV energies in the spectrum of sufficiently hard gamma-ray emitters. The best candidates for searches of GeV bumps are extreme blazars as their emitted flux is expected to show the largest TeV-to-GeV ratio among known extragalactic sources \cite{2020NatAs...4..124B}. The detection of the extreme blazar 1ES\:0229+200 ($z=0.14$ or $d\approx 650\:$Mpc) up to 10\:TeV \cite{2007A&A...475L...9A} triggered a series of joint GeV-TeV studies, with up-to-date reviews presented in Refs.~\cite{2021JCAP...02..048A, 2021Univ....7..223A, PueschelBiteau22}. After a decade of searches, no secondary GeV signal has been observed \cite{2018ApJS..237...32A}, be it as a spectral bump expected for a focused pair beam or as an angularly extended signal should the electrons and positrons be deflected by the IGMF. Current GeV-TeV constraints have thus been interpreted as placing a lower limit on the IGMF strength, at the level of $10^{-15}\:$G for a coherence length larger than 10\:kpc.\footnote{A recent study based on the stacking of gamma-ray images in the direction of radio galaxies claimed the exclusion of IGMF strengths larger than $10^{-15}\:$G \cite{2018ApJ...868...87B}, which is the complementary of the range tentatively excluded with conventional approaches. If confirmed, such an exclusion of the entire IGMF parameter space may rule out inverse-Compton cooling in favor of beam plasma instabilities, as initially proposed by these authors in Ref.~\cite{2012ApJ...752...22B}. The stacking constraints nonetheless rely on parametrizations of the luminosity function and gamma-ray spectrum of blazars and radio galaxies across the AGN unification scheme, whose present understanding appears to preclude a definite claim.} It should be noted that the limits imposed on the magnetic-field strength display order-of-magnitude variations when changing the source intrinsic parameters, in particular the duration of the active emission from the blazar, as well as the jet opening angle and orientation with respect to the line of sight. Despite being model-dependent, such gamma-ray constraints provide inputs to theoretical studies of the origin of cosmic magnetism \cite{2013A&ARv..21...62D, 2021RPPh...84g4901V}, which could arise either from astrophysical outflows or from a primordial phase transition. The IGMF strength expected from the latter scenarios is furthermore bound on the upper side by the CMB spectrum and anisotropies, at the level of $10^{-11}\:$G for the most recent constraints \cite{2019PhRvL.123b1301J}. More general limits, which are independent from the origin of cosmic magnetism, are placed on the maximum strength of the IGMF with rotation measures, at the level of $10^{-9}\:$G \cite{2016PhRvL.116s1302P}.

To conclude this section, while the pair-production process shown in the upper insert of Figure~\ref{fig:propagation} can nowadays be claimed as one of the observational discoveries of gamma-ray astronomy, the fate of the pairs in the lower insert remains elusive. No definite conclusion on the main dissipation process of electron-positron beams has been reached on theoretical grounds. The detection of a secondary gamma-ray signal at GeV energies would provide crucial inputs to IGMF models and could guide the identification of the seeds of cosmic magnetism in large-scale structures. 

\section{Deviations from the standard model at ultra-high and ultra-low masses}
\label{sec:pbsm}

\begin{figure}[ht!]
    \centering
    \includegraphics[width=1.22\linewidth,right]{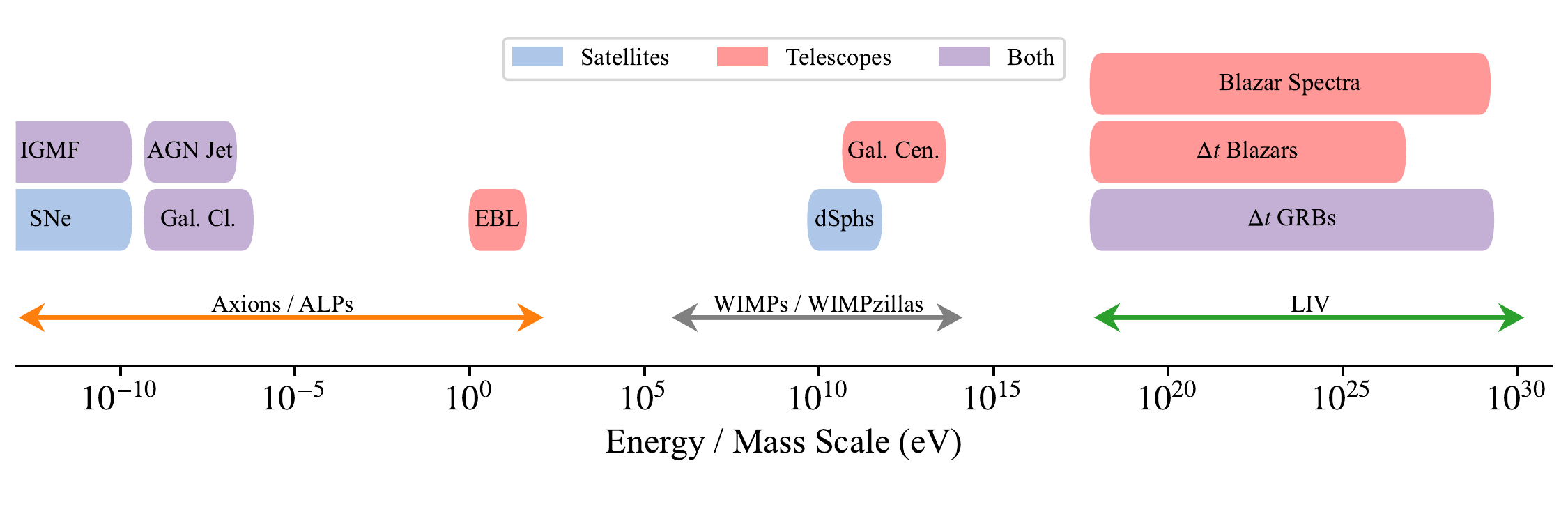}
    \caption{Energy and mass scales of physics beyond the standard model that can be probed with gamma-ray observations with satellites (blue boxes), ground-based telescopes (red boxes), or both (purple boxes). The range over which axions and ALPs are constrained lies below 10\:eV.
    The mass range for self-annihilating WIMPs, between $10^6$ and $10^{14}$\:eV, could be probed down to the thermal cross-section with future CTA observations of the Galactic center~(Gal. Cen.~\cite{2021JCAP...01..057A}) or \textit{Fermi}-LAT observations of dwarf spheroidal galaxies (dSphs \cite{2015PhRvL.115w1301A}). LIV is probed up to $10^{28}$\:eV ($n=1$) with both temporal and spectral observations as labelled in the Figure. The reader is referred to the repository in Ref.~\cite{ciaran_o_hare_2020_3932430} and Figure~1 in Ref.~\cite{2020Symm...12.1232M} for more specific constraints on the parameter spaces of ALPs and LIV, respectively.}
    \label{fig:pbsm}
\end{figure}

We have seen in Section~\ref{sec:journey} how processes beyond the standard model can alter cosmological gamma-ray propagation. 
These processes involve either the oscillation of photons in  axions and ALPs (with masses below $m_a \lesssim 10^{-6}\:\mathrm{eV}$) in astrophysical magnetic fields or LIV. The latter is expected at energy scales around the Planck energy $E_\mathrm{Pl}$.
In this section, we review how gamma-ray observations can put constraints on axions, ALPs, and LIV.
The constrained mass (or energy) scales are shown in Figure~\ref{fig:pbsm}.
For completeness, although not related to gamma-ray propagation, the masses of hypothetical weakly interacting massive particles (WIMPs) are also shown as they could self-annihilate (or decay) leading to the production of gamma-rays. We refer the reader to Ref.~\cite{2021arXiv211101198D} for a recent overview of how ground-based gamma-ray telescopes can probe WIMPs, ALPs, and other beyond-the-standard-model phenomena.

We first turn to the searches for axions and ALPs. 
The interest in these particles was spurred by a tentative observation with the PVLAS experiment of an optical rotation generated in vacuum by a magnetic field, which could be interpreted as evidence for ALPs~\cite{2006PhRvL..96k0406Z}.\footnote{The observation was ruled out a year later by the PVLAS collaboration after several experimental updates~\cite{2008PhRvD..77c2006Z}. See also Ref.~\cite{2020PhR...871....1E} for a recent review of the PVLAS experiment. It should be noted that photon-ALP oscillations in astrophysical environments had also been investigated prior to the PVLAS tentative signal. See, e.g., Refs.~\cite{1996PhRvL..77.2372G,1996PhLB..383..439B} for ALP-induced gamma-rays from the supernova  SN1987A, Refs.~\cite{2005PhRvD..72b3501M, 2005JCAP...02..005O} for type Ia supernovae dimming, and Ref.~\cite{2003JCAP...05..005C} for ultra-high-energy photons that could be observed beyond the threshold energy of pair production with CMB and radio photons.} 
It was soon realized that ALPs over a broad range of masses, $m_a$, and photon couplings, $g_{a\gamma}$,  would also affect the propagation of gamma-rays from extragalactic sources~\cite{2007PhRvD..76b3001M, 2007PhRvD..76l3011H}.
Since then, mainly two observational signatures have been identified: (a) spectral features---an energy dependent dimming of the flux in the form of (chaotic) oscillations---which occur around the critical energies given in Eqs.~\eqref{eq:ecrit-low} and \eqref{eq:ecrit-high}; (b) an enhancement of the gamma-ray flux in comparison to the case of pure absorption on background radiation fields. This enhancement should occur in the strong mixing regime, i.e.\ for gamma-ray energies $E_\mathrm{crit}^\mathrm{low} < E < E_\mathrm{crit}^\mathrm{high}$.

Various astrophysical magnetic fields along the line of sight have been studied in this context. 
Starting from the production site of gamma-rays within AGNs, 
several authors considered photon-ALP conversion in the magnetic field of the jet, the termination lobes of the jet and the host galaxy~\cite{2007PhRvL..99w1102H,2007PhRvD..76l3011H,2008PhRvD..77f3001S,2012PhRvD..86h5036T,2013JCAP...11..023M,2014JCAP...09..003M,2015PhLB..744..375T,2021PhRvD.103b3008D}.
In particular, a coherent toroidal component of the jet magnetic field, which can be of the order of $\mathcal{O}(\mathrm{G})$ at pc distances from the central black hole (e.g., Ref.~\cite{2014Natur.510..126Z}), could lead to spectral features in the gamma-ray range, as can be inferred from Eq.~\eqref{eq:ecrit-high}. 
The region of ALP masses that can be probed through the conversion in these magnetic fields is roughly $1\:\mathrm{neV}\lesssim m_a \lesssim 100\:\mathrm{neV}$
(region labeled ``AGN jet'' in Figure~\ref{fig:pbsm}). 
However, the exact morphology of the jet magnetic fields is still unknown, as is the location of the gamma-ray production site along the jet. Ideally these parameters are constrained through multi-wavelength observations and / or are left as additional nuisance parameters in an ALP hypothesis test~\cite{2021PhRvD.103b3008D}.
Radiation fields could also have a significant contribution to $E_\mathrm{crit}^\mathrm{high}$ in particular in FSRQs~\cite{2021DaviesMeyerCotter}.

Following the line of sight of the photon-ALP beam, the turbulent magnetic fields of galaxy clusters can act as efficient photon-ALP converters~\cite{2012JCAP...02..033H,2014JCAP...09..003M}. 
The fields could reach tens of $\mu\mathrm{G}$ in cluster centers~\cite{2012A&ARv..20...54F} and induce chaotic oscillations in gamma-ray spectra around $E_\mathrm{crit}^\mathrm{low}$. Observations with ground-based telescopes and \textit{Fermi}-LAT of bright gamma-ray emitting AGN in galaxy clusters provide world-leading constraints on the photon-ALP coupling for ALP masses around $1\:\mathrm{neV}\lesssim m_a \lesssim 1\:\mu\mathrm{eV}$ (the region labeled ``Gal. Cl.'', for Galaxy Clusters, in Figure~\ref{fig:pbsm}~\cite{2013PhRvD..88j2003A,2016PhRvL.116p1101A,2018PhRvD..97f3009Z,2018arXiv180504388M,2021JCAP...11..036G,2021PhLB..82136611C}).  
One of the most studied sources in this regard is NGC\:1275 at the center of the Perseus cluster. 
It should be noted though that the exact strength and configuration of the regular components of the large-scale magnetic field could change the ALP bounds considerably~\cite{2020PhLB..80235252L}.

After leaving the source and a potential galaxy cluster, photon-ALP conversion could also occur in the IGMF (see Section~\ref{sec:probing-ig} for a discussion of gamma-ray constraints on this magnetic field). 
This scenario has the appeal that the IGMF should be ubiquitously present in cosmic voids and, thus, one could expect photon-ALP conversion signatures in all AGN spectra. 
Using a cell-like turbulent IGMF with a field strength of the order of 1\:nG, which is close to the maximum value allowed by rotation measures~\cite{2016PhRvL.116s1302P}, 
it was realized that photon-ALP mixing could lead to an increased transparency of the Universe~\cite{2007PhRvD..76l1301D,2008PhLB..659..847D,2009JCAP...12..004M,2009MNRAS.394L..21D}.
A more realistic IGMF modeling, based on large-scale cosmological simulations of the magnetic field evolution, led to the same result~\cite{2017PhRvL.119j1101M,2017JCAP...01..024K,2018JHEAp..20....1G}.\footnote{As pointed out in Ref.~\cite{2014arXiv1412.2492M}, a cell-like model has the peculiarity that $P_{a\gamma}$ can be zero in the strong mixing regime for a certain alignment of the magnetic field.}
Interestingly, evidence for such a reduced opacity has been suggested by several authors and interpreted as evidence for ALPs~\cite{2007PhRvD..76l1301D,2009PhRvD..79l3511S,2011PhRvD..84j5030D,2012JCAP...02..033H,2011JCAP...11..020D,2013arXiv1310.2058M,2014JETPL.100..355R,2017PhRvD..96e1701K,2020MNRAS.493.1553G}.
This conclusion was reached mainly by observing an increasing gamma-ray flux in blazars at high energies and redshifts (and correspondingly large values of $\tau_{\gamma\gamma}$) when correcting the spectra for EBL absorption, which is not expected in standard blazar emission scenarios.
Lower limits on the photon-ALP coupling to explain these evidences were derived in Ref.~\cite{2013PhRvD..87c5027M}.
In contrast, other studies have confirmed that standard EBL absorption  describes the observed spectra well when large samples of spectra from ground-based telescopes and satellites are used~\cite{2015ApJ...812...60B,2015ApJ...813L..34D} and when the response of the instrument, in particular the energy resolution, is taken into account~\cite{2013A&A...554A..75S}.
Furthermore, a recent analysis of the highest energy photons recorded by the \textit{Fermi}-LAT showed no preference for the scenario of photon-ALP mixing in the IGMF~\cite{2020JCAP...09..027B}. 

Lastly, the photon-ALP beam enters the Milky Way. 
Its magnetic field can be modeled as a coherent and turbulent component (e.g., Ref.~\cite{2012ApJ...757...14J}). 
The coherence length for the turbulent component is too small compared to $\lambda_\mathrm{osc}$ to cause any significant oscillations. 
In contrast, depending on the line of sight, the coherent field can be of the order of $\mathcal{O}(\mu\mathrm{G})$ over kpc length scales leading to a potentially large oscillation probability~\cite{1996PhLB..383..439B,1996PhRvL..77.2372G,2008PhRvD..77f3001S,2012JCAP...02..033H}. 
The line-of-sight dependence through the Galactic magnetic field (GMF) could lead to a correlated detection of distant blazars (as ALPs reconvert back in the GMF to gamma-rays)~\cite{2014JCAP...01..016W} and an anisotropy in the diffuse gamma-ray background~\cite{2017arXiv171201839V,2021arXiv211003679C}. 
Furthermore, several authors have searched for ALP-induced spectral features around $E_\mathrm{crit}^\mathrm{low}$ in Galactic sources~\cite{2018JCAP...04..048M,2018PhRvD..97f3003X,2019PhRvD.100l3004X}. A preference for the existence of ALPs was for example claimed in the analysis of pulsar spectra measured with the \textit{Fermi}-LAT~\cite{2018JCAP...04..048M}. Such a claim is, however, at odds with results from the CAST experiment, which searches for ALPs produced in the Sun~\cite{2017NatPh..13..584A}.\footnote{The CAST bound could be avoided in scenarios discussed in Ref.~\cite{2021JCAP...11..036G}.}
Interestingly, the photon-ALP conversion in the GMF could also enable the detection of ALP-induced bursts of gamma-rays from core-collapse supernovae, which should arrive simultaneously with the neutrinos produced in the collapse.  
The non-detection of such a burst from SN\:1987A has lead to stringent bounds on ALPs with masses below $m_a \lesssim 1\:$neV~\cite{1996PhLB..383..439B,1996PhRvL..77.2372G,2015JCAP...02..006P}. ALP production in the supernova population has similarly been used to place constraints on the photon-ALP coupling~\cite{2020PhRvD.102l3005C}. These constraints would disfavor the ALP-induced reduction of gamma-ray opacity discussed above.\footnote{See also Ref.~\cite{2017arXiv171206205B} for a debate on the efficiency of ALP production during the core collapse.} 

Heavier axions and ALPs could also be probed indirectly by gamma-ray observations. 
The decay time of axions and ALPs into two photons is (e.g., Ref.~\cite{2008LNP...741...51R}) $\tau_{a\gamma\gamma} = 64\pi g_{a\gamma}^{-2} m_a^{-3} \approx 1.3 \ \times 10^{27}\:\mathrm{s}\ (g_{a\gamma} / 10^{-11}\mathrm{GeV}^{-1})^{-2} (m_a / \mathrm{eV})^{-3} $, which is ten orders of magnitude larger than the present age of the Universe. Decay of light ALPs discussed so far is thus irrelevant to observational probes. 
However, dark-matter axions and ALPs with masses around $1\:\mathrm{eV}\lesssim m_a\lesssim10\:\mathrm{eV}$ could decay into optical and UV photons and contribute to the EBL photon density (region labelled ``EBL'' in Figure~\ref{fig:pbsm})~\cite{2012JCAP...02..032C}. 
This would in turn increase the opacity of the Universe.
First analyses indicate that ground-based gamma-ray observations are promising to constrain these comparatively heavy ALPs~\cite{2019PhRvD..99b3002K,2020JCAP...03..064K,2020A&A...633A..74K}. 

Besides constraining the ALP parameter space, the propagation of gamma-rays provides constraints on LIV (see, e.g., Refs.~\cite{2020Symm...12.1232M,2021Univ....7..345T,PueschelBiteau22} for recent reviews). Similarly to hints for a reduced opacity discussed above for ALPs, the study of modifications of the pair-production threshold induced by LIV gained traction when the first blazars, namely Mrk\:421 and Mrk\:501, were observed in flaring states at energies up to $\sim20\:\mathrm{TeV}$~(e.g., Refs.~\cite{1999ApJ...518L..21K, 2001APh....16...97S} for historical references). Contrarily to ALP spectral modifications, the LIV effect is expected above a fixed energy, around $20\:$TeV for $n=1$ and $M = E_\mathrm{Pl}$. Such an effect emerges from the modified dispersion relation, to order $n$, of the particles involved in pair production, which has been treated either by deriving the modified threshold energy \cite{2008PhRvD..78l4010J} or by considering an effective mass for photons \cite{2014JCAP...06..005F}. Both approaches are purely based on kinematics and assume no modification of the pair-production cross section, which can only be predicted in specific extensions of the standard model that break Lorentz invariance (see discussion in Ref.~\cite{2016A&A...585A..25T} and references therein). 

Using state-of-the-art EBL models, there is currently no indication for a reduction of the Universe transparency to gamma-rays at very-high energies. Ground-based gamma-ray observations led to exclusions using searches in single observations~\cite{2019ApJ...870...93A} or combined analysis of multiple sources~\cite{2015ApJ...812...60B,2019PhRvD..99d3015L}, where the latter ones provide the strongest constraints up to $1-10\: E_\mathrm{Pl}$ (the probed range of $M$ is labeled ``Blazar spectra'' in Figure~\ref{fig:pbsm}). Interestingly, constraints on the LIV energy scale could also be derived from the non-observation of ultra-high-energy photons at $E>10^{18}\:\mathrm{eV}$ with the Pierre Auger Observatory~\cite{2018ApJ...853...23G}. 
For an unaltered pair-production threshold, a photon flux at these energies would be strongly absorbed in the interaction with CMB and radio photons. Not unlike gamma-ray constraints on the IGMF, ultra-high-energy constraints on LIV depend on the source properties which remain a limiting unknown.

The other channel through which gamma-ray propagation can be used as a probe of LIV lies in the study of time delays as a function of energy and distance~\cite{1998Natur.393..763A}. Ground-based observations of blazar flares have led to constraints on the LIV scale (box labeled ``$\Delta t$ Blazars'' in Figure~\ref{fig:pbsm}) to be above $0.2\: E_\mathrm{Pl}$~\cite{2008PhLB..668..253M,2011APh....34..738H,2019ApJ...870...93A}. Ground-based gamma-ray telescopes have also recently discovered TeV emission from GRBs~\cite{2019Natur.575..455M,2021Sci...372.1081H,2021ApJ...908...90A}. In particular, the detection by MAGIC of GRB~190114C ($z = 0.42$) at energies above $0.2\:\mathrm{TeV}$ ruled out $M < 0.5 \: E_\mathrm{Pl}$~\cite{2020PhRvL.125b1301A}. The most stringent time-delay constraints come from \textit{Fermi}-LAT observation of GRB\:090510 at a redshift of $z\sim0.90$ which rules out $M < 2\: E_\mathrm{Pl}$~\cite{2013PhRvD..87l2001V}. The values of $M$ probed with GRBs are shown in Figure~\ref{fig:pbsm} in the region labelled ``$\Delta t$ GRBs''.
It should be noted that even though these time-delay analyses achieve strong constraints on the LIV scale $M$, assumptions have to made about the intrinsic time spread of the emission. This problem is starting to be addressed both for AGN flares~\cite{2020A&A...633A.143P} and GRBs~\cite{2020ApJ...890..169P,2021ApJ...906....8D} in a systematic way, but further work appears to be needed to disentangle intrinsic from possible line-of-sight effects~\cite{2021JCAP...05..029A}. 

\section{Knowns, expectations and hopes from gamma-ray cosmology}

The exploration of the extragalactic gamma-ray sky around TeV energies has unveiled some of the most extreme accelerators in the Universe: from blazars detected up to 20\:TeV at light travel times of ${\sim}\:500\:$Myrs ($z\approx 0.03$) out to distant ones detected above 200\:GeV at ${\sim}\:8\:$Gyrs ($z\approx 1$). Gamma-ray observations of jetted AGNs at very-high energies provide powerful probes of the content of intergalactic medium, as discussed in Section~\ref{sec:probing-ig}, and of physics beyond the standard model, as discussed in Section~\ref{sec:pbsm}. The recent discovery of emission beyond 100\:GeV from long GRBs also provides enticing prospects to complement the explorations performed with AGNs.

The long-sought signal of gamma-ray absorption on the EBL, predicted in the 1960s, has been uncovered by observations from both gamma-ray satellites and ground-based telescopes. Measurements of the EBL imprint thus far primarily constrain the present-day COB at optical and near-infrared wavelengths, with a precision better than $30\%$. Gamma-ray astronomy at intermediate and high redshifts ($z>1$) from space has further provided its first competitive constraints on the EBL evolution with redshift, which is dictated by the cosmic history of star formation, dust consumption and AGN formation. The study of the baryonic and light contents of the ionized Universe, and of their co-evolution, is expected to strongly benefit from continued operation of \textit{Fermi}-LAT in space during the early-science phase of the new-generation gamma-ray observatory, the Cherenkov Telescope Array (CTA \cite{2019scta.book.....C}). Through its key-science programs, CTA will be able to probe the EBL evolution out to at least $z=2$ with a precision two-to-three times better than current-generation telescopes \cite{2021JCAP...02..048A}. CTA will in particular probe with unprecedented accuracy the diffuse or unresolved emissions in the region of the ``optical controversy'', in conjunction with deep- and wide-field observations from JWST, the Rubin Observatory, \textit{Euclid} and the Roman Space Telescope \cite{2019scta.book.....C}. The possible identification of distant (re-)ionizing sources with dedicated optical and near-infrared observations together with CTA constraints on emission from the overall population(s) can let us hope for major steps forward in constraining the low-end of the galaxy luminosity function and in the understanding of the intergalactic medium. At higher EBL wavelengths, around $100\:\mu$m, CTA observations of high emission states from nearby extragalactic sources are expected to place tight constraints on the CIB, not only assessing the cosmic-dust emissivity but also resolving an important source of uncertainties in the propagation of ultra-high-energy nuclei ($E > 10^{18}\:$eV) as they photodissociate on this infrared background \cite{2015JCAP...10..063A}. As ultra-high-energy nuclei are charged, their propagation is also affected by magnetic fields on galactic, intrahalo, intracluster and intergalactic scales, which could impact the interpretation of anisotropies detected on large-angular scales at $E > 8 \times 10^{18}\:$eV \cite{2017Sci...357.1266P} and suggested on intermediate angular scales at $E > 40 \times 10^{18}\:$eV \cite{2018ApJ...853L..29A}. New insights on cosmic magnetism are expected from radio observations of clusters and filaments with SKA and LOFAR  (see, e.g., the MAGCOW project \cite{2021arXiv211109129V}). Such observations could guide numerical simulations of large-scale structures and inform gamma-ray searches for IGMF signatures. CTA in particular offers the potential to probe the IGMF in voids up to strengths of at least 0.3\:pG \cite{2021JCAP...02..048A}. A strong interplay of theoretical/numerical advances and astroparticle/astronomical observations offers exciting prospects in our understanding of the intergalactic medium in the upcoming decade.

As we have seen in Section~\ref{sec:pbsm}, searches for physics beyond the model affecting gamma-ray propagation have also become a vibrant field of study. CTA will in particular have unprecedented sensitivity searching for spectral irregularities ~\cite{2021JCAP...02..048A} and a reduced gamma-ray opacity (e.g., Refs.~\cite{2014JCAP...12..016M,2017JCAP...01..024K}) as induced by ALPs. With the foreseen energy range of CTA, it will be possible to probe higher ALP masses potentially up to hundreds of neV for couplings $\gtrsim10^{-12}\:\mathrm{GeV}^{-1}$ for cluster magnetic fields of the order of $10\:\mu\mathrm{G}$ \cite{2021JCAP...02..048A}.
At these masses and couplings, CTA observations will begin to probe the parameter space where ALPs could constitute the entire cold dark matter content of the Universe. 
A similar region of the ALP parameter space could be probed across the energy ranges of the TAIGA detector~\cite{2020JInst..15C9031B} and the LHAASO water Cherenkov detector~\cite{2019arXiv190502773B}.  
For high values of $g_{a\gamma}$ (or in optimistic magnetic field scenarios), LHAASO could further detect reconverted ALPs at the high-energy end of the diffuse gamma-ray background~\cite{2021JCAP...11..030L}.
Space-based gamma-ray observatories could also detect ALP-induced bursts of gamma-rays from supernovae, either in the Milky Way~\cite{2017PhRvL.119j1101M} or in close-by galaxies~\cite{2020PhRvL.124w1101M}. Additionally, theoretical advancements open up the parameter space of the QCD axion towards lower masses which could be probed with gamma-ray instruments (e.g., Refs.~\cite{2017JHEP...01..095F,2021JHEP...05..184D,2021JHEP...06..123S}). 
The development of open-source software~\cite{2021arXiv210802061M,2021ascl.soft09001M} facilitates the computation of photon-ALP oscillation probabilities in various magnetic field environments for young scientists entering the domain. 
Possible evidences or constraints derived from future gamma-ray observations should be easily testable with the next-generation of axion experiments, also coming online in the next decade (e.g., Ref.~\cite{2018PrPNP.102...89I}).
In that sense, the search for axions and ALPs with gamma-ray observations nicely complements the experimental efforts to search for this kind of particle.  

Searches for LIV will greatly benefit from deep-field observations with CTA as well as from wide-field observations beyond tens of TeV with LHAASO and SWGO. In light of the recent transient events detected with already operating ground-based telescopes, detections of GRBs with CTA appear to be guaranteed, presumably at a rate larger than $0.1-1$ per year that was initially suggested \cite{2012MNRAS.425..514K,2013ExA....35..413G,2013APh....43..252I}. While LHAASO and SWGO will continuously monitor a large fraction of the sky and possibly detect some serendipitous events \cite{2020ApJ...900...67K}, CTA will have to follow-up on external triggers provided by multi-wavelength facilities such as \textit{Swift}, SVOM or \textit{Fermi}-GBM~\cite{2019scta.book.....C}.  For the linear subluminal LIV scenario, the sensitivity of CTA could reach $10\:E_\mathrm{Pl}$ depending on the time delay between the GRB trigger and the start of the observation~\cite{2013APh....43..252I}. 
LHAASO could even reach a sensitivity to LIV of the order of $\sim20\:E_\mathrm{Pl}$~\cite{2019arXiv190502773B}. 
These limits will of course depend on a better understanding of the intrinsic time lags, which need to be incorporated in the data analysis~\cite{2021ApJ...906....8D,2021JCAP...05..029A}. 
Furthermore, optical follow-up observations will be of paramount importance in order to determine the redshift and hence the distance of GRBs. 
CTA observations of blazars should also improve considerably the sensitivity to alterations of the pair-production threshold. 
The authors of Ref.~\cite{2021JCAP...02..048A} used simulated CTA observations of a flare from Mkn\:501 as well as long-term observations of the hard-spectrum blazar 1ES\:0229+200 to estimate the possible constraints on $M$ in the linear and quadratic LIV scenario.
With single observations of these sources, energy scales $M \sim E_\mathrm{Pl}$ could be achieved in the linear case and a source stacking should push the sensitivity a factor two-to-three beyond current constraints. 

To conclude, the study of gamma-ray propagation has placed competitive constraints on cosmological entities and on fundamental physics over the past decade. Such constraints and discoveries have been enabled by the capablities of instruments such as \textit{Fermi}-LAT, H.E.S.S, MAGIC and VERITAS, which have left us with challenging questions: \textit{Have we resolved only half of the optical light in the Universe?} \textit{Where is cosmic magnetism stemming from?} \textit{Could dark matter consist in light particles such as axions?} \textit{Is Lorentz symmetry broken beyond the Planck scale?} We look forward to CTA and other upcoming gamma-ray observatories to help the multi-wavelength and multi-messenger communities take up the challenge.

\authorcontributions{J.~B.\ and M.~M.\ have equally contributed, read and agreed to the published version of the manuscript. Data Curation, Writing, Visualization: Sections~1 and 3 by J.~B., Sections~2 and 4 by M.~M., Section~5 by both.}

\funding{M.~M.\  acknowledges  support from the European Research Council (ERC) under the European Union’s Horizon 2020 research and innovation program Grant agreement No. 948689 (AxionDM) and from the Deutsche Forschungsgemeinschaft (DFG, German Research Foundation) under Germany’s Excellence Strategy – EXC 2121 ``Quantum Universe''.}

\dataavailability{Data used in the graphs presented in this review can be requested from the authors.} 

\acknowledgments{J.~B. and M.~M. would like to thank the guest editors John Quinn, Deirdre Horan and Elisa Pueschel for proposing a special issue of Galaxies on Extragalactic TeV Astronomy. J.~B.\ acknowledges exchanges with Humberto Mart\'inez-Huerta on LIV. J.~B. and M.~M. are grateful to the reviewers for their feedback on the manuscript.}

\conflictsofinterest{The authors declare no conflict of interest.} 

\abbreviations{Abbreviations}{
The following abbreviations are used in this manuscript:\\

\noindent 
\begin{supertabular}{ll}
1EG	        &	First EGRET gamma-ray catalog\\
1FGL	    &	First \textit{Fermi}-LAT catalog\\
1FLGC	    &	First \textit{Fermi}-LAT gamma-ray burst catalog\\
1FLT	    &	First \textit{Fermi}-LAT catalog of long-term gamma-ray transient sources\\
2FLGC	    &	Second \textit{Fermi}-LAT gamma-ray burst catalog\\
3EG	        &	Third EGRET catalog of high-energy gamma-ray sources\\
3FGL	    &	Third \textit{Fermi}-LAT catalog\\
4FGL-DR2	&	Fourth \textit{Fermi}-LAT catalog, data release 2\\
AGN	        &	Active galactic nucleus\\
AGU     	&	Active galactic nucleus of uncertain type (\textit{source type})\\
ALP	        &	Axion-like particle\\
AMEGO	    &	All-sky medium energy gamma-ray observatory (\textit{instrument})\\
BCU 	    &	Blazar of uncertain type (\textit{source type})\\
BH	        &	Black hole\\
BLL	        &	BL\:Lac source (\textit{source type})\\
BLR	        &	Broad-line region\\
CAST	    &	CERN axion solar telescope (\textit{instrument})\\
CAT	        &	Cherenkov array at Themis (\textit{instrument})\\
CGRO	    &	Compton gamma-ray observatory (\textit{instrument})\\
CIB	        &	Cosmic infrared background, from mid to far infrared\\
CMB	        &	Cosmic microwave background\\
COB	        &	Cosmic optical background, from near ultraviolet to near infrared\\
CPT         &   Charge, parity and time-reversal symmetries\\
CSFH	    &	Cosmic star-formation history\\
CSS	        &	Compact steep spectrum source (\textit{source type})\\
CTA	        &	Cherenkov Telescope Array (\textit{instrument})\\
DEVILS	    &	Deep extragalactic visible legacy survey\\
dSph        &   Dwarf spheroidal galaxy\\
e-ASTROGAM	&	Enhanced ASTROGAM mission (\textit{instrument})\\
EBL	        &	Extragalactic background light, from near ultraviolet to far infrared\\
EGRET	    &	Energetic gamma ray experiment telescope (\textit{instrument})\\
EW	        &	Electroweak\\
\textit{Fermi}-GBM	&	\textit{Fermi} gamma-ray burst monitor (\textit{instrument})\\
\textit{Fermi}-LAT	&	\textit{Fermi} large area telescope (\textit{instrument})\\
FSRQ	    &	Flat-spectrum radio quasar (\textit{source type})\\
GAL	        &	Normal galaxy (\textit{source type})\\
Gal. Cen.    &	Galactic center\\
Gal. Cl.    &	Galaxy cluster\\
GAMA	    &	Galaxy and mass assembly survey\\
GMF	        &	Galactic magnetic field\\
GRB	        &	Gamma-ray burst (\textit{source type})\\
H.E.S.S.	&   High energy stereoscopic system	(\textit{instrument})\\
HAWC        &	High-altitude water Cherenkov observatory (\textit{instrument})\\
HEGRA	    &	High-energy-gamma-ray astronomy (\textit{instrument})\\
ICRC 	    &	International cosmic-ray conference\\
IGL	        &	Integrated galaxy light\\
IGMF	    &	Intergalactic magnetic field\\
JWST	    &	James Webb Space Telescope (\textit{instrument})\\
lGRB	    &	Long gamma-ray burst (\textit{source type})\\
LHAASO 	    &	Large high altitude air shower observatory (\textit{instrument}) \\
LI	        &	Lorentz invariance\\
LIV	        &	Lorentz-invariance violation\\
LOFAR 	    &	Low-Frequency Array (\textit{instrument}) \\
MAGCOW	    &	Magnetized cosmic web project\\
MAGIC	    &	Major atmospheric gamma imaging Cherenkov telescopes (\textit{instrument})\\
NLSY1	    &	Narrow-line Seyfert 1 galaxy (\textit{source type})\\
PVLAS 	    &	Polarizzazione del vuoto con laser (\textit{instrument})\\
QCD	        &	Quantum chromodynamics\\
RDG	        &	Radio galaxy  (\textit{source type})\\
SBG	        &	Starburst galaxy  (\textit{source type})\\
SEY	        &	Seyfert galaxy  (\textit{source type})\\
sGRB	    &	Short gamma-ray burst (\textit{source type}) \\
SKA 	    &	Square Kilometer Array (\textit{instrument}) \\
SN  	    &	Supernova (\textit{source type}) \\
SSRQ	    &	Steep-spectrum radio quasar (\textit{source type})\\
SVOM        &   Space variable objects monitor (\textit{instrument}) \\
TAIGA	    &	Tunka advanced instrument for cosmic ray physics and gamma astronomy (\textit{instrument})\\
TeVCat	    &	TeV catalogue\\
UV	        &	Ultraviolet\\
VERITAS	    &	Very energetic radiation imaging telescope array system (\textit{instrument})\\
WIMP	    &	Weakly interacting massive particle\\
$\Lambda$CDM	& Lambda cold dark matter model\\
\end{supertabular}}

\appendixtitles{no} 

\end{paracol}
\reftitle{References}

\newpage
\externalbibliography{yes}
\newcommand{\apj}{ApJ}
\newcommand{\apjs}{ApJS}
\newcommand{\apjf}{The Astrophysical Journal}
\newcommand{\apjl}{ApJL}
\newcommand{\apjlf}{The Astrophysical Journal Letters}
\newcommand{\aap}{A\&A}
\newcommand{\aapf}{Astronomy \& Astrophysics}
\newcommand{\aapr}{A\&A~Rev.}
\newcommand{\apss}{Ap\&SS}
\newcommand{\aaprf}{Astronomy \& Astrophysics Reviews}
\newcommand{\araa}{Annual Rev. of Astron. and Astrophys.}
\newcommand{\araaf}{Annual Review of Astronomy and Astrophysics}
\newcommand{\epjc}{Eur. Phys. J. C}
\newcommand{\epjcf}{European Physical Journal C}
\newcommand{\epjwc}{Eur. Phys. J. Web Conf.}
\newcommand{\epjwcf}{European Physical Journal Web of Conferences}
\newcommand{\nat}{Nature}
\newcommand{\science}{Science}
\newcommand{\ap}{Astrop. Phys. }
\newcommand{\apf}{Astroparticle Physics Journal }
\newcommand{\ijmpd}{Int. J. of Mod. Phys. D}
\newcommand{\ijmpdf}{International Journal of Modern Physics D}
\newcommand{\jcap}{J. Cosmology Astropart. Phys.}
\newcommand{\jcapf}{Journal of Cosmology and Astroparticle Physics}
\newcommand{\jhea}{J. High-Energy Astroph.}
\newcommand{\jheaf}{Journal of High Energy Astrophysics}
\newcommand{\jpcs}{J. Phys. Conf. Ser.}
\newcommand{\jpcsf}{Journal Physics Conference Series}
\newcommand{\mnras}{MNRAS}
\newcommand{\mnrasf}{Monthly Notices of the Royal Astronomical Society}
\newcommand{\nima}{Nucl. Instrum. and Meth. in Phys. Research A}
\newcommand{\nimaf}{Nuclear Instruments and Methods in Physics Research A}
\newcommand{\plb}{Phys. Let. B}
\newcommand{\plbf}{Physics Letter B}
\newcommand{\prd}{Phys.~Rev.~D}
\newcommand{\prdf}{Physical Review D}
\newcommand{\prl}{Phys.~Rev.~Letter}
\newcommand{\prlf}{Physical Review Letter}
\newcommand{\physrep}{Physics Reports}
\newcommand{\pasa}{Publications of the Astron. Soc. of Australia}
\newcommand{\pasj}{Pub. of the ASJ}
\newcommand{\pasjf}{Publications of the Astronomical Society of Japan}
\newcommand{\pos}{PoS}
\newcommand{\posf}{Proc. of Science}
\newcommand{\rpp}{Rep. on Prog. in Phys.}
\newcommand{\rppf}{Reports on Progress in Physics}
\newcommand{\ssr}{S Review}
\newcommand{\etal}{\emph{et al}}
\newcommand{\coll}{Collaboration}
\newcommand{\baas}{Bulletin of the American Astron. Soc.}
\bibliography{my_external_BibTeX_file}

\end{document}